\documentclass[sigplan,10pt]{acmart}
\renewcommand\footnotetextcopyrightpermission[1]{}
\newcommand{\vny}[1]{\textcolor{blue}{vny: #1}}

\AtBeginDocument{%
  \providecommand\BibTeX{{%
    \normalfont B\kern-0.5em{\scshape i\kern-0.25em b}\kern-0.8em\TeX}}}

\graphicspath{{./images/}}

\usepackage{tikz}
\usetikzlibrary{positioning, arrows, backgrounds, patterns, calc, shapes, fit}
\usepackage{listings}
\usepackage{xcolor}
\usepackage{tcolorbox}
\usepackage{colortbl}
\usepackage{graphicx}
\tcbuselibrary{listings,skins}
\usepackage{algorithm}
\usepackage{algpseudocode}   

\usepackage{amsmath}
\usepackage{amssymb}
\usepackage{multirow}
\usepackage{booktabs}
\usepackage{scalerel}
\usepackage{booktabs}
\usepackage[subtle]{savetrees}
\usepackage{hyphenat}
\usepackage{enumitem}
\usepackage{subcaption}
\setlist[itemize]{align=left, leftmargin=*}
\setlist[enumerate]{align=left, leftmargin=*}

\definecolor{initcode}{RGB}{213,232,212}    
\definecolor{cleancode}{RGB}{255,242,204}   
\definecolor{guideaccess}{RGB}{225,213,231} 
\definecolor{initoutline}{RGB}{130,179,102} 
\definecolor{cleanoutline}{RGB}{214,182,86} 
\definecolor{guideoutline}{RGB}{150,115,166} 

\newcommand*\circledBlue[1]{\tikz[baseline=(char.base)]{
            \node[shape=circle,fill={rgb,255:red,46; green,111; blue,158},inner sep=1.3pt] (char) {\textcolor{white}{#1}};}}            

\newcounter{finding}
\newcommand{\finding}[1]{\addvspace{0.2em}\noindent\fbox{\parbox{\columnwidth}{\textit{\textbf{Finding \stepcounter{finding}\thefinding:} #1}}}\addvspace{0.2em}}

\newcommand{\SYSTEM}{\textit{OBASE}}
\newcommand{\SYSTEMfull}{\textit{O}bject-\textit{B}ased \textit{A}ddress-\textit{S}pace \textit{E}ngineering}

\newcommand{\ARC}{ATC}
\newcommand{\ARCfull}{Active Thread Count}
\newcommand{\OC}{OC}

\newcommand{\TAG}{TAG}
\newcommand{\TAGfull}{Thread-local Active scope Guard}
\newcommand{\TAI}{TAI}
\newcommand{\TAIfull}{Thread Activity Index}

\newcommand{\CIW}{CIW}
\newcommand{\CIWfull}{Consecutive Inactive Window}
\newcommand{\Ct}{$C_t$}

\newcommand{\SODA}{SODA}

\newcommand{\ODMfull}{Optimistic Data Migration}
\newcommand{\ODM}{ODM}
\newcommand{\PRactual}{$PR_{actual}$}
\newcommand{\PRtarget}{$PR_{target}$}

\lstdefinestyle{threadcode}{
  basicstyle=\scriptsize\ttfamily,
  columns=flexible,
  keepspaces=true,
  basewidth=0.5em,
  resetmargins=true,
  xleftmargin=0pt,
  xrightmargin=0pt,
  aboveskip=0pt,
  belowskip=0pt,
  lineskip=-2pt,
  boxpos=t,
  framesep=0pt
}

\newcommand{\myparagraph}[1]{%
  \noindent
  \textbf{
    #1%
    .
  }
}

\begin{document}

\newcommand{\PlainAuthors}{Vinay Banakar, Suli Yang, Kan Wu, Andrea C. Arpaci-Dusseau, Remzi H. Arpaci-Dusseau, Kimberly Keeton}
\renewcommand{\thefootnote}{\fnsymbol{footnote}}

\author{
  Vinay Banakar\textsuperscript{1, 3}, 
  Suli Yang\textsuperscript{3}, 
  Kan Wu\textsuperscript{2}\footnotemark[1],
  Andrea C. Arpaci-Dusseau\textsuperscript{1}, \\ 
  Remzi H. Arpaci-Dusseau\textsuperscript{1}, 
  Kimberly Keeton\textsuperscript{3} \vspace{1.5ex} \\ 
  \textsuperscript{1}University of Wisconsin-Madison \quad 
  \textsuperscript{2}xAI \quad 
  \textsuperscript{3}Google \\
}


\title {\texttt{OBASE:} Object-Based Address-Space \\Engineering to Improve Memory Tiering}

\begin{abstract}
Hardware and OS mechanisms for memory tiering are widely deployed, yet datacenters still overprovision DRAM.
The root cause is \textit{hotness fragmentation}: allocators place objects by size rather than access pattern, so hot and cold objects become interleaved within the same pages.
A single hot object marks its page as active, trapping surrounding cold data in expensive DRAM.
Our analysis of Google datacenter workloads shows that up to 97\% of the bytes in active pages are cold and unreclaimable.
We propose \textit{address-space engineering}: dynamically reorganizing virtual memory so that hot objects cluster into uniformly hot pages and cold objects into uniformly cold pages.
We present \SYSTEM{}, a compiler-runtime system for unmanaged languages that serves as an object-aware \textit{frontend} for page-aware OS \textit{backends}.
\SYSTEM{} tracks accesses via lightweight pointer instrumentation and migrates objects at runtime using a lock-free protocol that is safe under concurrency.
By reorganizing the address space, \SYSTEM{} enables unmodified backends (kswapd, TMO, TPP, Memtis) to tier memory effectively.
Across ten concurrent data structures, six backends, and production traces from Meta and Twitter, \SYSTEM{} improves page utilization by 2--4$\times$ and reduces memory footprint by up to 70\%, with only 2--5\% overhead.

\end{abstract}

 \settopmatter{printacmref=false}

\maketitle
\thispagestyle{empty}

\footnotetext[1]{Work done at Google}

\pagestyle{empty}

\section{Introduction}

Memory capacity, especially DRAM, has become the dominant cost driver in modern data centers, often accounting for 50\% of server capital expenses~\cite{google-borg, tpp, azure-mem-cost}. Memory tiering promises a solution: by placing cold, less accessed data in cheaper, slower tiers—such as compressed memory, SSDs, or CXL-attached disaggregated memory~\cite{cxl, intel-optane-pmm, ull, samsung-memory-ssd, zswap, cxl-on-azure}—one could significantly reduce the expensive DRAM capacity required, thereby lowering costs~\cite{TMO, tpp, memtis, autonuma, tmts, hemem}.

\begin{figure}[!t]
    \centering
    \includegraphics[width=0.49\textwidth]{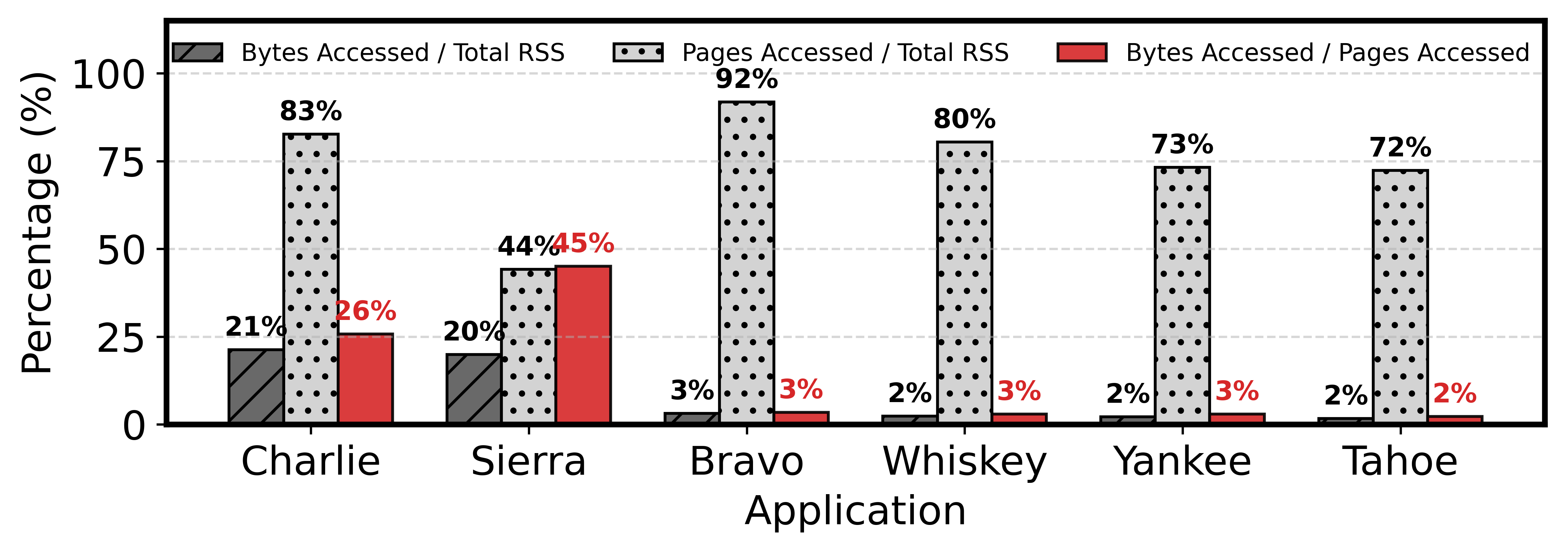}
    \caption{
        \textbf{The granularity gap in memory access of Google workloads.} 
        \small{Dark Grey: The percentage of total memory \textit{bytes} actually accessed. Light Grey: The percentage of total memory \textit{pages} accessed. Red: The page utilization (calculated as Bytes Accessed / Pages Accessed, see \S\ref{sec:quantify-fragmentation} for more details).}
    }
    \label{fig:google-itntro}
\end{figure}

In principle, this approach shows great promise, primarily because in hyperscale workloads, only a small volume of memory is hot; the vast majority remains cold and un-accessed for extended periods. As shown in Figure~\ref{fig:google-itntro}, across six different Google workloads~\cite{Google_Workload_Traces_Version_2}, only 1.7\%--21.3\% of bytes are accessed during the trace duration. Indeed, four of the six workloads access less than 3\% of bytes, suggesting that we could reclaim 80\% to 98\% of DRAM for secondary tiers with minimal impact on application performance.

In practice, however, realizable gains often fall short of this potential. Google reported offloading only 20\% of infrequently accessed data to a compressed memory tier~\cite{zswap}, while Meta achieved similarly modest savings of 20--32\%~\cite{TMO}.

The crux of this gap lies in a mismatch between how applications organize data and how the OS manages memory: applications access data at \textit{object} granularity, which varies in size, while the OS manages memory at fixed-size \textit{page} granularity (4KB, 2MB, or 1GB)~\cite{ArpaciDusseau23-Book}. Modern allocators place objects on pages with no regard for how these objects will be accessed in the future, often placing frequently accessed objects and rarely touched objects intermingled on the same page. As a result, a few hot bytes (e.g., a single object) would make a whole page active from the OS's perspective, trapping the whole page in expensive DRAM. One can clearly see this phenomenon in Figure~\ref{fig:google-itntro}, where even though only 3.2\% of \textit{bytes} are accessed in a workload Bravo, 91.8\% of the \textit{pages} are accessed, making them unfit candidates for reclamation. Similar trends can be observed across all six workloads. 
We term this phenomenon \textbf{hotness fragmentation} and identify it as the root cause of inefficient tiering at hyperscalers. We quantify this fragmentation using the \textit{page utilization} metric.


To combat hotness fragmentation, we introduce \textbf{address space engineering}, which \textit{dynamically} reorganizes objects within the address space so that objects with similar access intensities are grouped together. By segregating hot and cold objects, address space engineering promises to close the gap between theoretical and realizable memory savings. It creates a memory layout with uniformly cold pages (or hot) that are significantly more amenable to tiering.

\begin{figure*}[!t]
\centering
\includegraphics[width=0.98\textwidth]{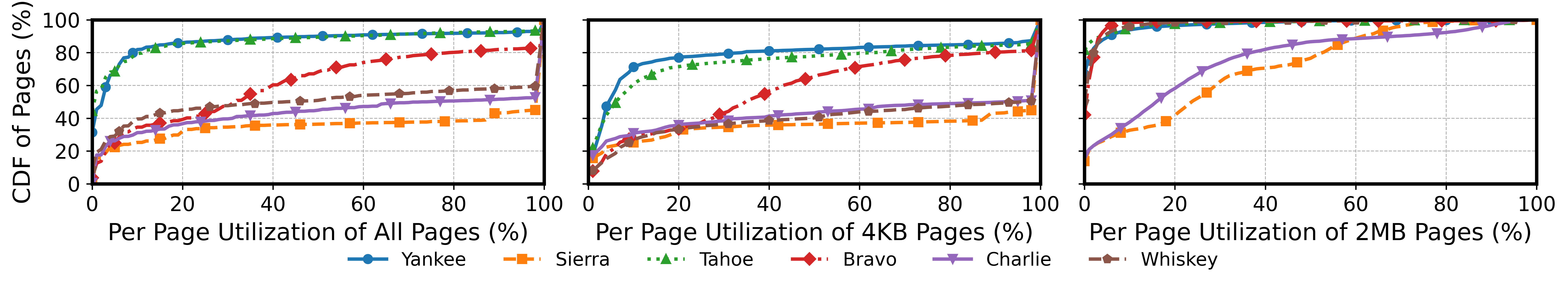}
\vspace{-4mm}
\caption{
    \textbf{Low Per-Page Utilization in Google Data Center Workloads.} 
    \small{CDFs of page utilization for six applications, shown for all pages combined (left) and separately for 4KB pages (center) and 2MB pages (right). 
    }
}
\label{fig:google-rio-pu}
\end{figure*}

Crucially, by focusing on the address space \textit{layout}, we decouple memory tiering into two orthogonal concerns, allowing independent innovation within each: the \textit{layout} (frontend) problem, which organizes objects to yield high-quality page candidates for reclamation; and the \textit{reclamation} (backend) problem, which focuses on migration policies and mechanisms across different tiers. This separation allows us to reuse existing page-based tiering infrastructures—both swap-based~\cite{zswap, TMO} and byte-addressable~\cite{tpp, memtis, hemem}—and leverage future improvements in reclamation mechanisms. It also ensures that our layout optimization techniques remain applicable to future, currently non-existent memory tiers (e.g., CXL 3.0 fabrics or NVM). Further, it significantly lowers the adoption barrier for hyperscalers: rather than requiring changes to their existing tiering backends, address space engineering provides a superior memory layout that makes existing backends work more effectively. 

In this paper, we present \textbf{\SYSTEMfull{} (\SYSTEM)}, a compiler-runtime system that engineers the address space for \textit{unmanaged} languages like C/C++. \SYSTEM{} acts as an intelligent \textit{frontend} for memory organization: it manages pointer-based data structures, transparently profiles object access using lightweight instrumentation to determine access intensity, and employs a novel lock-free protocol to migrate objects safely, clustering hot and cold objects together. As a result, \SYSTEM{} prepares the memory layout for any OS \textit{backend}, enabling them to work more effectively. \SYSTEM{} preserves a familiar programming model: developers annotate which pointer fields may relocate, and the compiler routes their accesses through guides. This does require giving up two assumptions (stable object addresses and pointer arithmetic over managed objects) and applies to pointer-based structures rather than arbitrary code; we make this boundary explicit in \S\ref{sec:limitations}.

We evaluate \SYSTEM{} with ten concurrent pointer-based data structures and six state-of-the-art reclamation and tiering backends~\cite{kswapd, tpp, memtis, autonuma, TMO}. Using key-value store workloads (YCSB) and production traces (Meta, Twitter), we demonstrate that \SYSTEM{} improves page utilization (by 2--4$\times$) and achieves higher memory savings (up to 70\%) at the same performance overhead, or conversely, achieves equivalent savings with negligible performance impact. 
For tiered-memory configurations, \SYSTEM{} achieves the same throughput with half the DRAM capacity.



\section{Case for Dynamic Object Reorganization}
\label{sec:motivation}

In this section, we show that hotness fragmentation is both severe and unavoidable under static object placement. 
We introduce a metric to quantify fragmentation, analyze data center traces from Google~\cite{Google_Workload_Traces_Version_2}, and demonstrate that object hotness changes continuously over time. 
Finally, we explain why unmanaged languages make dynamic reorganization especially challenging, motivating the design of \SYSTEM{}.

\subsection{Page Utilization: A Metric}
\label{sec:motivation-utilization}

The OS marks a page as ``active’’ if it receives at least one access during a time window $T$. 
We define \textit{per-page utilization} as the fraction of a touched page’s capacity that is actually accessed. 
Let $P(T)$ be the set of touched pages during $T$, and let $U(p,T)$ be the number of unique bytes accessed within page $p$ during $T$. 
We then define the aggregate \textit{page utilization} as:

\vspace{-2mm}
\[
\text{Page Utilization}(T) = 
\frac{\sum_{p \in P(T)} U(p,T)}
     {\sum_{p \in P(T)} \text{Size}(p)}
\]
\vspace{-2mm}

Low utilization means a page appears hot to the OS even when most of its bytes are cold.  
Previous work~\cite{tidy-up} demonstrated low page utilization for open-source databases (e.g., Redis, MongoDB) where 75--90\% of accessed pages utilized less than 15\% of their capacity for YCSB workloads.

\subsection{Fragmentation in Data Center Workloads}
\label{sec:quantify-fragmentation}

To understand hotness fragmentation across a broader range of applications, we analyze memory access traces from six data center workloads at Google~\cite{Google_Workload_Traces_Version_2}, collected using Dynamo\-RIO~\cite{DynamoRIO}.  We process each unique access, annotate pages as 4KB or 2MB using \texttt{/proc/self/smaps}, and count unique 64B cache lines touched per page. Based on instruction counts, core counts, and typical warehouse-scale CPU characteristics~\cite{profile-wsc}, we estimate the traces capture up to $\sim$30s of steady-state execution.

\begin{figure*}[!t]
    \centering
    \begin{subfigure}[b]{0.495\textwidth}
        \includegraphics[width=\textwidth]
        {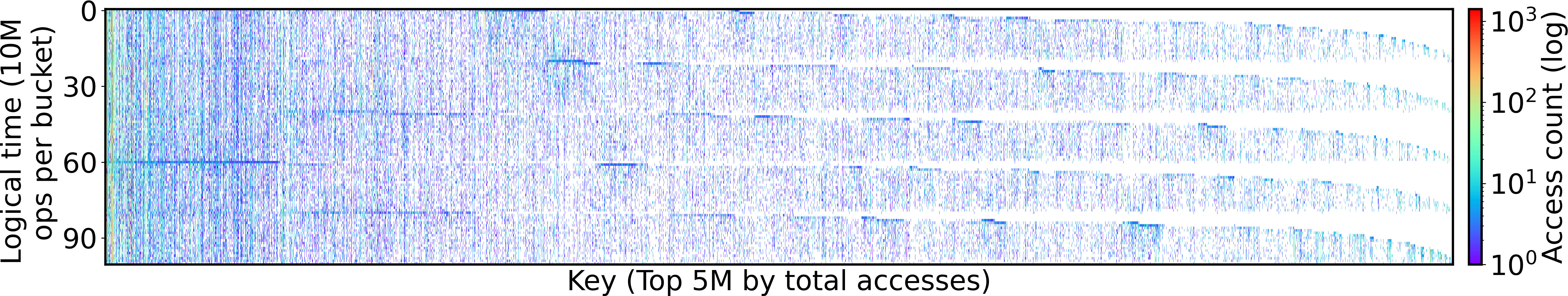}
        \caption{\textbf{Meta KV Trace.} \small{White horizontal bands indicate coordinated quiet periods where access drops across many keys. These bands reveal phased workload behavior.}}
        \label{fig:heatmap-meta}
    \end{subfigure}
    \hfill
    \begin{subfigure}[b]{0.495\textwidth}
        \includegraphics[width=\textwidth]
        {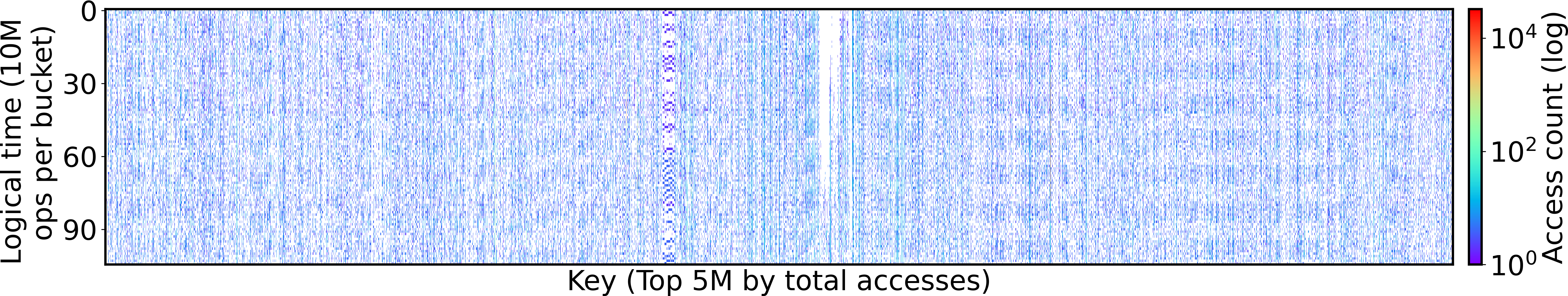}
        \caption{\textbf{Twitter Trace.} 
        \small{A sparse, scattered pattern indicates sporadic hotness. A few keys on the far left remain consistently hot, but the majority exhibit bursty access with long idle gaps.}}
        \label{fig:heatmap-twitter}
    \end{subfigure}
    \vspace{-3mm}
    \caption{\textbf{Temporal Evolution of Key Hotness.} \small{Heatmaps showing access frequency (log scale) for the top 5M keys over logical time buckets (10M operations each). If hotness were static, we would observe continuous vertical bands on the left side of each plot. Instead, we see shifting phases (Meta) and intermittent bursts (Twitter), demonstrating that the hot working set evolves continuously.}}
    \label{fig:heatmaps}
\end{figure*}

Figure~\ref{fig:google-rio-pu} shows the cumulative distribution of per-page utilization for all workloads, ordered as: (a) aggregated across both 4KiB and 2MiB page sizes, (b) 4KB pages only, and (c) 2MB pages only.


\textbf{Aggregated view.} Aggregated across both page sizes, all six workloads show populations of poorly-utilized pages. Yankee and Tahoe are the most fragmented: median utilization is around 3\%, meaning half of their touched pages waste over 95\% of capacity on cold data. Bravo follows with a median near 35\%. The remaining workloads have higher overall utilization but still carry long left tails: roughly 25--30\% of pages in Sierra, Charlie, and Whiskey use less than 20\% of their capacity. Page-based reclamation cannot distinguish these poorly-utilized pages from well-utilized ones without object-level information.

\textbf{4KB pages.}
Fragmentation is visible even at the smallest page size: for Tahoe and Yankee $\sim$80\% of pages fall below 20\% utilization, for Bravo 60\% fall below 40\%, and Charlie, Whiskey, and Sierra have 35--40\% of pages below 40\%. Hotness fragmentation is thus inherent to access patterns, not merely an artifact of huge pages.

\textbf{2MB pages.}
Fragmentation worsens dramatically for huge pages. Tahoe, Bravo, and Yankee are extreme: 85--90\% of their huge pages utilize under 10\% of their 2MB capacity, and even Sierra has 40\% below 20\%. Each such page, consuming 512$\times$ a base page, holds over 1.8MB of cold data alongside a few accessed cache lines. \\
\finding{Active pages are mostly cold: 70--90\% of bytes in pages the OS considers hot receive no accesses.}

\subsection{Object Hotness Changes Over Time}

\label{sec:motivation-dynamic}

One potential solution for hotness fragmentation is to segregate hot and cold objects at allocation time, using static hints or profiling \cite{llvm-pgho}.  
This strategy assumes that hotness is both identifiable at allocation time and relatively stable afterward.

First, identifying hotness at allocation time is challenging because 
the same code path allocates objects with vastly different lifecycles.  For example, in Redis and Memcached, a single \texttt{SET} handler allocates memory for all incoming records, yet one record might be a session token accessed every millisecond, while another is a user profile never touched again. Static analysis cannot distinguish these cases at allocation time.  Second, this approach assumes that object hotness is relatively stable: objects identified as hot when allocated will remain hot throughout their lifetimes. 

We show that these assumptions do not hold. We analyze object-level traces from Meta~\cite{cachelib-osdi} and Twitter~\cite{twitter-traces}, which record operations in logical time (10M operations per bucket).  Figure~\ref{fig:heatmaps} visualizes access intensity for the top 5 million keys (ranked by total accesses) over 100 such buckets. If hotness were static, the most popular keys (left) would exhibit continuous vertical bands.  Instead, both traces show significant churn: bursts of activity followed by long idle gaps.

\textbf{Meta: Phased hotness.} 
The Meta workload shows coordinated phases where many keys become inactive at once, visible as white horizontal bands. Even keys that are repeatedly accessed overall alternate between active bursts and extended dormancy.  A page packed with currently-hot objects eventually becomes a page dominated by cold objects as access patterns shift.

\textbf{Twitter: Sporadic hotness.} 
The Twitter workload shows a sparse pattern. 
A small slice of keys remains consistently hot, but the majority exhibit brief access bursts separated by long idle gaps, even among the top 5 million keys.

To quantify churn, we measure reuse-distance variability (75th to 25th percentile of number of operations between accesses to the same key).  
For mid-sized objects (64B--4KB), representing 94\% of Meta keys and 98.2\% of Twitter keys, 75\% of keys have a reuse spread exceeding $5\times$, and 65\% exceed $30\times$.  Thus, access gaps for the majority of keys fluctuate by more than an order of magnitude.  \\

\finding{Hotness is transient; object hotness is neither knowable at allocation time nor stable over time. As a result, one-time placement cannot prevent hotness fragmentation, and dynamic object migration based on changing hotness is required.}

\subsection{Object Mobility in Unmanaged Languages}
\label{sec:principle_mobility}

Moving objects dynamically presents varying degrees of difficulty depending on the language runtime. Managed runtimes, such as the JVM and Go, can relocate objects during garbage collection by updating references atomically, making mobility relatively straightforward. In unmanaged languages like C++, however, programs assume that an object's address is stable~\cite{mythGC}, rendering dynamic object migration significantly more challenging.
Thus, the challenge is not merely to reorganize the layout dynamically, but to do so without breaking program correctness under aliasing and concurrency.

n this paper, we demonstrate how to enable object mobility in \textit{unmanaged} languages specifically because of this challenge. We show that address-space engineering is achievable in C++, where stable addresses are the default assumption, by routing accesses through an indirection rather than preserving the address itself; techniques proven in this setting generalize naturally to managed runtimes.

To make the problem tractable in an unmanaged context, we focus on \textbf{pointer-based data structures}, where elements are accessed indirectly through pointers. These structures are ubiquitous in memory-intensive applications (Table~\ref{tab:ds}). Rather than attempting to intercept arbitrary raw-pointer dereferences, which is intractable in C++, we route accesses to relocatable objects through an explicit indirection object (a \emph{guide}, \S\ref{sec:progmodel}) that the application uses in place of a raw pointer. When an object moves, the system redirects the guide to its new location transparently. This mechanism must operate safely in highly concurrent environments, without resorting to coarse-grained locks or stop-the-world pauses that would negate the performance benefits of an improved memory layout. 

In \S~\ref{sec:design}, we show how \SYSTEM{} provides this transparency and concurrency within the C++ execution model, enabling dynamic reorganization.

\section{\SYSTEMfull{}}
\label{sec:design}

\begin{figure}[t]
  \centering
  \includegraphics[width=0.48\textwidth]{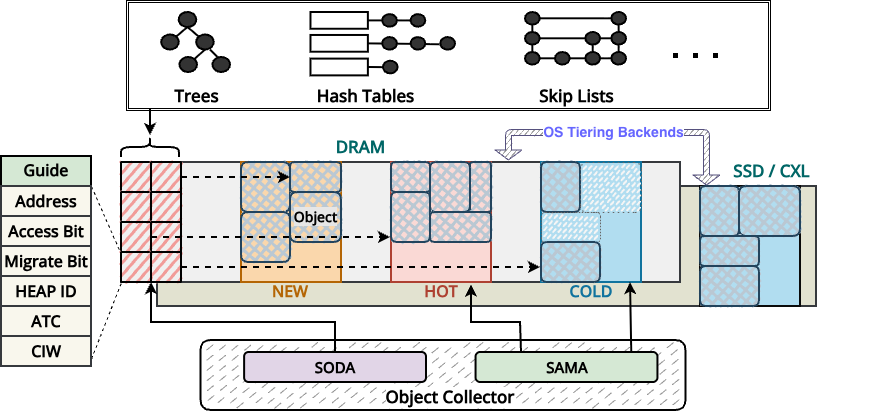}
  \caption{\textbf{\SYSTEM{} Overview.} 
  \small{\SYSTEM{} acts as a frontend that organizes the virtual address space into hot and cold regions. The Object Collector monitors access via Guides and SODA, migrates objects using SAMA, and presents a re-organized address space to the OS backend.}}
  \label{fig:hades-overview}
  \vspace{-3mm}
\end{figure}


Our goal is to engineer an application's address space so that page-based backends can efficiently reclaim or tier memory.  
\SYSTEM{} achieves this by reshaping object placement: cluster hot objects into dense regions to increase page utilization, and segregate cold objects into separate regions to expose large pools of reclaimable memory.

\SYSTEM{} operates in environments where the operating system manages memory in page-sized units and may migrate pages across tiers~\cite{ArpaciDusseau23-Book, ull, zswap, demystifying-cxl}.
The design does not assume a particular tiering policy or memory hierarchy; it only assumes that backends observe per-page activity.
By presenting backends with regions that are uniformly hot or cold, 
\SYSTEM{} allows existing mechanisms—from traditional page reclaim (kswapd, zswap) to tiering engines (such as TMO, TPP, Memtis, and HeMem)—to make better decisions without becoming object-aware~\cite{TMO, zswap, tpp, memtis, hemem}.

Achieving object-level placement with page-level backends poses four challenges: \textbf{(C1)~object mobility}---C++ ties object identity to its address, so relocation must be transparent and safe under aliasing; \textbf{(C2)~low-overhead tracking}---hotness classification must run in the common path without degrading performance; \textbf{(C3)~dynamic adaptation}---hotness shifts continuously (\S\ref{sec:motivation-dynamic}), so the layout must evolve; and \textbf{(C4)~safe concurrent migration}---objects must move without global pauses even while threads hold pointers to them.

\subsection{System Overview}
\label{sec:overview}


\SYSTEM{}'s design rests on separating the \emph{layout} problem (organizing objects so pages contain uniformly hot or cold data) from the \emph{tiering} problem (deciding which pages to evict and where to put them). We call systems addressing the layout problem \emph{frontends} and those addressing reclamation \emph{backends}: existing tiering systems like Kswapd, TPP, Memtis, and TMO are backends that assume the layout is given. \SYSTEM{} is a frontend that takes responsibility for layout so any backend can operate on uniformly hot or cold pages. 

The following subsections describe how \SYSTEM{} enables object mobility (\S\ref{sec:progmodel}), tracks accesses efficiently (\S\ref{sec:tracking}), organizes the address space (\S\ref{sec:address-space-org}), and migrates objects safely under concurrency (\S\ref{sec:saftey}).

Figure~\ref{fig:hades-overview} illustrates the architecture. \SYSTEM{} runs in a continuous control loop.  All dereferences of managed objects pass through a lightweight {\it guide} abstraction, which records whether an object was accessed recently.
A background \textit{Object Collector (OC)}, periodically processes this metadata to classify objects by access activity and decide whether they should reside in the NEW, HOT, or COLD heaps. 
Based on this classification, the OC reorganizes the address space by migrating objects between heaps using a safe, lock-free protocol based on \textit{Active Thread Counts (ATC)}.
Finally, \SYSTEM{} exposes these organized regions to the OS through a \textit{Spatially-Aware Memory Allocator (SAMA)}, which lays out each heap as a contiguous virtual range, enabling coarse-grained OS hints.
This control loop ensures that the virtual address space continuously adapts to the application's shifting working set while presenting page-based backends with clear hot and cold targets.

\subsection{Object Mobility in Unmanaged Languages}

\label{sec:progmodel}


\SYSTEM{} decouples an object’s logical identity from its location using a lightweight \emph{guide} abstraction. A guide carries the current location of the object as well as additional metadata needed by \SYSTEM{}.  Developers interact with objects by using guides rather than raw pointers.  When a guide is dereferenced, \SYSTEM{} resolves it to the object’s current address and records that the object was accessed (described in~\S\ref{sec:tracking}). The indirection layer is the mechanism that makes later relocation transparent; code that previously operated on pointers continues to operate on guides.

Developers choose which pointers can participate in relocation by marking them with annotations (e.g., the pointers to keys and values in hash-table buckets or the record pointers in B+ trees).  
In practice this annotation is small: for the ten data structures in our evaluation (Table~\ref{tab:ds}), each managed type carries one to three relocatable pointer fields (typically the structure's authoritative child) while container algorithms and client code are untouched.
Guides are enforced through three compiler passes (detailed in \S\ref{sec:llvm-pass}). A type-level pass identifies annotated pointers and rewrites their dereferences to invoke the guide. An instrumentation pass injects hooks at access sites so \SYSTEM{} can observe uses without modifying application logic. A validation pass ensures that managed objects are not used in unsupported ways (e.g., pointer arithmetic over nodes or assumptions about physical contiguity). These passes allow developers to adopt \SYSTEM{} incrementally, starting with the portions of a codebase where object residency matters most.

The division of labor is deliberately narrow: the \emph{developer} marks a small set of pointer fields and guarantees that no unannotated raw pointer aliases a managed object; the \emph{compiler} performs everything else: rewriting declarations and dereferences, and inserting the tracking hooks. An annotated guide carries unique ownership of its object, analogous to \texttt{std::unique\_ptr}: \SYSTEM{} updates that single guide when relocating, and does not track hidden aliases, so structures that share a node through multiple pointers (e.g., graphs, doubly-linked lists) are out of scope. The validation pass rejects pointer arithmetic and physical contiguity assumptions over managed objects rather than silently miscompiling them. A raw pointer obtained from a guide dereference (e.g., held in a register) remains valid for the duration of the enclosing public operation, because that operation keeps the object's active-thread count above zero and thereby vetoes any concurrent migration (\S\ref{sec:saftey}); callers must not cache such pointers beyond the operation.

Two further programming-model restrictions keep these invariants intact under concurrent migration. First, the \texttt{Guide<T>} type appears only inside the data structure's implementation, where Pass 2 rewrites the developer-marked pointer fields (\S\ref{sec:llvm-pass}); it does not appear in calling code. Second, guides do not cross the data structure's public boundary as parameters or return values; they are held as internal fields, and callers interact through primitive-typed APIs (e.g., \texttt{HashMap::get(int idx)} in Figure~\ref{fig:code-transformation}). Together these ensure that every guide dereference is reachable from a public-method entry whose TAG participates in the migration protocol.

\subsection{Low-Overhead Access Tracking}
\label{sec:tracking}

To classify objects by temperature, \SYSTEM{} must observe which objects are accessed over time, but tracking must be cheap enough to run continuously. Existing mechanisms fall short at both ends of the spectrum. Hardware page table access bits operate at page granularity and cannot distinguish a few hot cache lines from megabytes of cold data on the same page~\cite{damon, zswap}. Software profilers such as DynamoRIO and LLVM's memory profiling provide fine-grained information but impose prohibitive overheads for always-on deployment~\cite{llvm-memprof, DynamoRIO}. Hardware sampling via PEBS offers lower overhead but provides statistical coverage rather than precise per-object tracking~\cite{pebs}.

Instead, \SYSTEM{} embeds access tracking directly into guide pointer dereferences, yielding object-level information without significant overhead.  A guide overloads the dereference operator so that each access records that the underlying object was used. Modern 64-bit architectures reserve high-order address bits in canonical user-space pointers, so \SYSTEM{} stores a small amount of per-object metadata in these unused bits~\cite{tagged-ptr}. Inline metadata avoids external side tables and ensures that recording access is part of normal pointer use.

Metadata is updated on every dereference using a single atomic read–modify–write (RMW): an \texttt{accessed} bit is set if it has not already been set, thereby skipping subsequent updates to avoid unnecessary cache-coherence traffic for frequently touched objects. This design keeps the common path small, and the resulting per-access cost is comparable to a cache hit. The metadata also holds state used by the runtime to classify objects or coordinate relocation, without requiring separate allocations or indirection. 

To let the runtime observe all managed objects without walking application-specific pointer graphs, \SYSTEM{} maintains a Sparse Object Data Activity (\SODA{}) bitmap (\S\ref{sec:impl-soda}) over the process heap. SODA records which virtual regions contain managed objects and enables the Object Collector (\OC{}) to discover objects by scanning these regions rather than interpreting container internals.

\subsection{Dynamically Engineering the Layout}
\label{sec:address-space-org}

The access signals from \S\ref{sec:tracking} feed into a layout policy that continuously reorganizes the virtual address space. OBASE groups objects by observed temperature so that page-based backends see large, uniform regions rather than interleaved hot and cold data. Three logical temperature heaps capture this temperature: NEW holds freshly allocated objects whose access pattern is not yet clear, HOT holds the current working set, and COLD holds objects inactive for multiple scan periods. Each heap occupies a dedicated virtual address range, so an object's address directly encodes its temperature.

\begin{figure}[t]
  \centering
  \includegraphics[width=0.46\textwidth]{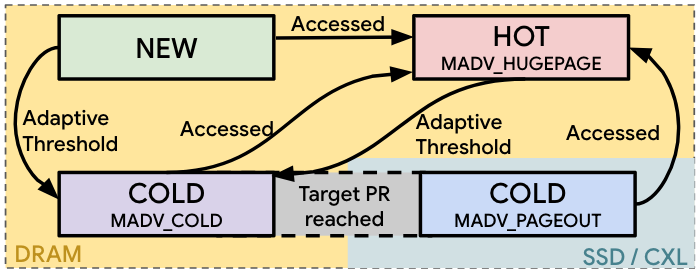}
  \vspace{-2mm}
  \caption{\textbf{Object Migration State Diagram.} 
  \small{Objects transition between heaps based on access intensity. The Object Collector promotes accessed objects to HOT and demotes inactive objects to COLD, allowing SAMA to apply different \texttt{madvise} policies to each region if required.}}
  \label{fig:heap-state-diagram}
\end{figure}

Objects move between heaps as access intensity changes (Figure~\ref{fig:heap-state-diagram}): inactive HOT objects are demoted to COLD, and COLD objects that become active again are promoted back. This continuous reclassification tracks the workload's evolving working set rather than its allocation history.

To realize these logical heaps in the virtual address space, \SYSTEM{} employs a Spatially-Aware Memory Allocator (SAMA) that reserves a contiguous virtual address range for each heap and sub-allocates objects within that range. Contiguity is a deliberate design choice that allows coarse-grained OS hints to be applied to whole heaps rather than individual pages, exposing large pools of cold memory to page-based tiering systems. The Object Collector (OC) periodically scans SODA to classify each managed object. Objects touched since the last scan are candidates for promotion; objects that remain untouched accumulate evidence of coldness.

Reacting to a single inactive window would make classification sensitive to transient bursts. \SYSTEM{} therefore tracks a per-object \CIWfull{}(\CIW{}) counter: each scan window without an access increments CIW; any access resets it to zero. Objects whose CIW exceeds a cold threshold become eligible for demotion to COLD; objects in COLD that are accessed rejoin HOT. This hysteresis ensures that only sustained inactivity triggers migration to COLD.

The cold threshold \Ct{} governs how long an object must remain inactive before migration. Too aggressive and COLD objects are frequently re-accessed; too conservative and reclaimable memory lingers in HOT. \SYSTEM{} adapts \Ct{} using a promotion-rate target. We define the promotion rate (PR) as the fraction of the working set drawn from COLD heap in each scan window:
\[
PR = \dfrac{\text{unique COLD pages accessed}}{\text{working set size}} \times \dfrac{60}{\text{scan interval (s)}}
\]
Crucially, \SYSTEM{} measures COLD-heap accesses regardless of where those pages physically reside—it cannot determine which tier a page occupies and does not attempt to. If the observed rate exceeds a target, \Ct{} increases by one window; if below, \Ct{} decreases. This additive adjustment converges to a workload-specific regime. The goal is not to minimize COLD-heap accesses, but to ensure that pages classified as COLD are \textit{safe targets} for any backend policy. \S\ref{sec:implementation} details initialization and tuning.

By design, \SYSTEM{} is decoupled from any specific backend. The base system reorganizes the address space but issues no reclamation hints. This separation is intentional: the value of address-space-engineering lies in making pages uniformly hot or cold, which improves the precision of any page-based mechanism. Backends such as kswapd, TMO, TPP, and Memtis observe per-page activity through their existing interfaces (PTE-A bits, PEBS samples, or PSI signals) and naturally make better decisions when COLD pages contain only cold objects.

Optionally, \SYSTEM{} can issue proactive hints to accelerate reclamation. In this hinted mode, once the promotion rate stabilizes below the target, \SYSTEM{} marks COLD pages with \texttt{MADV\_COLD} or \texttt{MADV\_PAGEOUT} to signal that they are safe to reclaim. Similarly, SAMA may request huge pages for HOT (\texttt{MADV\_HUGEPAGE}) to selectively improve TLB coverage over dense hot data. These hints are strictly advisory and complement rather than replace backend policies.

\subsection{Safe Concurrent Migration}
\label{sec:saftey}

Reorganizing the address space requires relocating objects while application threads may hold pointers to them. In managed languages, garbage collectors solve this problem using load barriers or stop-the-world pauses. C++ offers neither: there is no runtime to intercept pointer loads, and production systems cannot tolerate pauses. OBASE must therefore migrate objects without blocking threads, while guaranteeing that every dereference sees a valid location. As described in \S\ref{sec:progmodel}, all dereferences pass through guides, so relocation reduces to atomically updating a single pointer-sized word. Callers never follow stale pointers and do not need explicit synchronization.

\begin{figure*}
\centering
\begin{minipage}[c]{0.20\textwidth}
\vstretch{1.18}{\includegraphics[width=\textwidth]{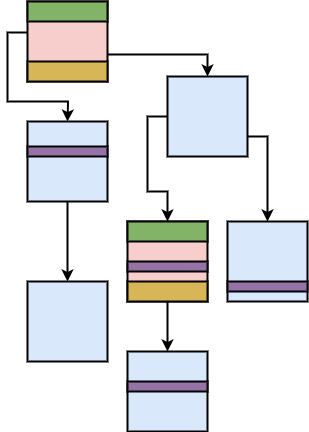}}
\end{minipage}
\begin{minipage}[c]{0.78\textwidth}

    \includegraphics[width=\textwidth]{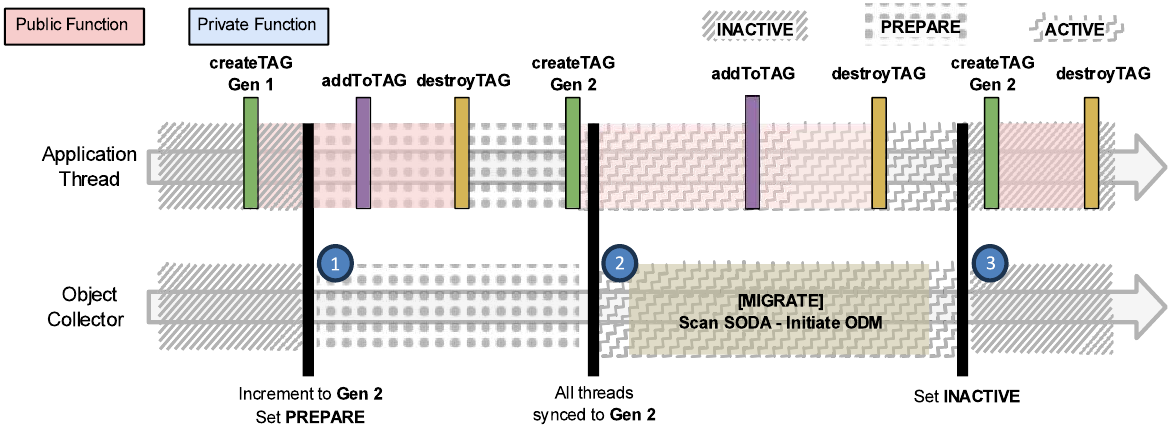}
  
  
  \begingroup
  \renewcommand{\arraystretch}{0.8}  
  \begin{tabular}{|>{\centering\arraybackslash}p{0.30\textwidth}|>{\centering\arraybackslash}p{0.30\textwidth}|>{\centering\arraybackslash}p{0.30\textwidth}|}
    \hline
    \cellcolor{initoutline} \textbf{\textcolor{white}{\texttt{\small Start of Public Function}}} & 
    \cellcolor{guideoutline} \textbf{\textcolor{white}{\texttt{\small Guide Access}}} & 
    \cellcolor{cleanoutline} \textbf{\textcolor{white}{\texttt{\small End of Public Function}}} \\
    \hline
    \cellcolor{initcode} \texttt{\scriptsize \textbf{createTAG}} &
    \cellcolor{guideaccess} \texttt{\scriptsize \textbf{addToTAG(Guide)}} &
    \cellcolor{cleancode} \texttt{\scriptsize \textbf{destroyTAG}} \\
    \hline
  \end{tabular}
  \endgroup
\end{minipage}
\caption{\textbf{Application Thread Execution Flow And Interaction With OC} 
\small{The left side shows the instrumented call graph, code is inserted in three scenarios as shown. Public functions create and destroy \TAGfull{}s (\TAG{}s), maintaining nesting levels and registering the thread in the \TAIfull{} (\TAI{}). Guides increment active reference counts only if added to the \TAG{}. Reference counts are decremented only when the outermost public function exits. \ARCfull{} (\ARC{}) is enabled only during \texttt{PREPARE} and \texttt{ACTIVE} states.}}
\label{fig:thread-flow}
\end{figure*}
   
             

\paragraph{Lightweight Scoping.}
An object can be migrated only when no thread is actively using it. 
Classic approaches—stop-the-world pauses or per-access load barriers—are unsuitable for C++ systems code~\cite{zgc, shenandoah, rcu}.
Instead,\SYSTEM{} introduces a per-object \emph{Active Thread Count} (ATC) that tracks how many public data structure operations have observed a guide during a migration window. The ATC captures the notion of active use: if a thread begins an operation that may dereference an object, ATC is incremented once for that scope; when the operation completes, the ATC is decremented. Migration is permitted only when the ATC reaches zero, indicating that no thread is currently executing an operation that could read or modify the object.


The ATC must be updated efficiently. Rather than incrementing the ATC on every pointer dereference—which would impose unacceptable overhead—\SYSTEM{} scopes tracking to public API boundaries. When a thread enters a data-structure operation (e.g., get, insert), it registers with a Thread-local Active scope guard (\TAG{}); when the operation completes, the guard decrements the ATC for all objects touched. This design reflects how C++ programmers reason about pointer validity: pointers are valid for the duration of the operation that obtained them, not indefinitely. Compiler instrumentation (detailed in \S\ref{sec:impl-tag}) inserts the necessary hooks automatically.

\paragraph{Active Windows via Epochs.}
Always-on ATC tracking would impose overhead even when no migration is in progress. \SYSTEM{} therefore activates ATC only during periodic migration epochs. Outside these windows, guide dereferences record only access activity (\S\ref{sec:tracking}) with no ATC overhead. When the OC initiates migration, it coordinates an epoch transition that ensures all active threads enable ATC tracking before any object is moved. This epoch-based activation confines the synchronization cost to brief, infrequent windows. 

\paragraph{Optimistic Migration.}
Given ATC=0, \SYSTEM{} moves objects using an optimistic protocol inspired by optimistic concurrency control. The Object Collector copies the object to its target heap, then attempts to atomically swing the guide to the new address. If any thread accesses the object during the copy, the commit fails and the move is abandoned—the object remains in place, and the thread sees valid data. This optimistic approach has two key properties: (1) threads never block on migration, and (2) concurrent access safely vetoes relocation rather than observing inconsistent state. \S\ref{sec:impl-epoch} details the CAS-based protocol.

\paragraph{Safety and Non-Blocking Progress.}
Safety follows from a single invariant: the guide is updated atomically, and migration aborts on any conflicting access. No thread can ever follow a stale pointer. Progress is non-blocking: application threads never wait on the collector, and the collector performs bounded work per object before committing or abandoning. Frequently accessed objects naturally resist migration (their ATC rarely reaches zero), while cold objects eventually move. The result is an address space that reshapes continuously without stop-the-world pauses, while preserving familiar pointer semantics.

\section{Implementation}
\label{sec:implementation}

This section details \SYSTEM{}'s concrete realization.\footnote{\SYSTEM{} is open-sourced at \url{https://github.com/WiscADSL/obase}} We cover the guide metadata encoding (\S\ref{sec:impl-guide}), scope-guard data structures (\S\ref{sec:impl-tag}), SODA bitmap layout (\S\ref{sec:impl-soda}), controller parameters (\S\ref{sec:impl-controller}), the epoch-based migration protocol (\S\ref{sec:impl-epoch}), kernel reclamation optimizations (\S\ref{sec:impl-kernel}), and the compiler passes that automate guide management (\S\ref{sec:llvm-pass}).


\subsection{Guide Metadata Encoding and Heap Allocation}
\label{sec:impl-guide}

Each guide uses the 48 bits required for canonical x86-64 user-space addresses and repurposes the upper 16 bits for metadata: 7 bits encode ATC (supporting up to 128 concurrent threads per object), 5 bits track CIW (up to 32 windows, or 60+ minutes at the default 120s interval), 2 bits identify the current heap (NEW, HOT, COLD), and 2 bits store the access and migration-lock flags. As 128-bit addressing becomes more prevalent~\cite{128b-linux}, implementations can expand these fields without changing the guide abstraction. We implement the three heaps using a Spatially-Aware Memory Allocator (SAMA) built on jemalloc's extent management. SAMA reserves large \texttt{mmap} regions for each heap and allocates objects within them, returning unused extents to the OS as objects migrate or are freed.

\subsection{Efficient Scope Guard Tracking}
\label{sec:impl-tag}

Figure~\ref{fig:thread-flow} illustrates the instrumented call graph.
The compiler inserts three hooks: \texttt{createTAG} (green) at public API entry,
\texttt{addToTAG(guide)} (purple) at each guide dereference, and \texttt{destroyTAG} (yellow) at public API exit.
Public functions may call private helpers that also dereference guides; the TAG tracks all such accesses but only decrements the ATC when the \emph{outermost} public function returns. This nesting-aware design ensures that the ATC remains positive throughout the entire operation, even when internal helpers are inlined or called multiple times.

We implement TAGs using a \texttt{BaseDeltaPtrSet} that exploits pointer locality:
it encodes pointers as a base address plus 32-bit deltas, grouping up to 16 nearby pointers within two cache lines. Insertion is O(1) when the pointer falls within an existing group; otherwise a new group is created in O(log G) time, where G is the number of groups. 
Since pointers within a single operation cluster tightly (e.g., keys and values in the same bucket, nodes along a tree path), most insertions hit existing groups. Across the ten data structures
in our evaluation, the median number of unique guides per operation ranges from 3 (hash tables) to 12 (B+Tree traversals), keeping per-operation TAG overhead under
100\,ns.

\subsection{SODA Bitmap Structure and Object Discovery}
\label{sec:impl-soda}
The Sparse Object Data Activity (SODA) bitmap uses a two-level structure to cover the heap address space without allocating storage for empty regions. SODA divides the address space into coarse-grained blocks; each block contains 64-bit words where individual bits indicate guide presence at fixed-size slots. Blocks are allocated lazily and reclaimed when empty. Because SODA tracks guide pointers rather than object addresses, it remains valid across relocations; the guide's address does not change when the object moves. Memory overhead is one bit per potential guide slot plus block-level bookkeeping. The OC scans SODA at a configurable interval (120\,s by default), chosen to align with cold-memory detection thresholds in warehouse-scale tiering systems~\cite{tmts, zswap}.

\subsection{Cold Threshold Controller}
\label{sec:impl-controller}
\SYSTEM{} initializes \Ct{} to three scan windows; at the default 120\,s interval, this is approximately six minutes of inactivity before demotion, consistent with the five-minute rule for tiered storage~\cite{five-min-recent, five-min-ten-later, five-min-og}. The target promotion rate of 1\% is based on budgets used in production compressed-memory and CXL tiering deployments~\cite{zswap, tmts}. \Ct{} is bounded between 1 and 32 windows.
After each scan window, an additive-increase/additive-decrease (AIAD) loop (Algorithm~\ref{alg:controller}) compares the observed promotion rate $PR_{actual}$ (\S\ref{sec:address-space-org}) against the target and adjusts \Ct{} accordingly, so the threshold settles into a workload-specific regime without large swings.

\begin{algorithm}[t]
\footnotesize
\caption{Cold-Threshold Controller (per scan window)}
\label{alg:controller}
\begin{algorithmic}[1]
\State $PR_{actual} \leftarrow \dfrac{\text{unique COLD pages accessed}}{\text{working-set size}} \times \dfrac{60}{\text{scanInterval}}$
\If{$PR_{actual} > PR_{target}$} \Comment{demotions revisited too often}
    \State $C_t \leftarrow \min(C_t + 1,\ 32)$ \Comment{demote more conservatively}
\ElsIf{$PR_{actual} < PR_{target}$} \Comment{slack in the budget}
    \State $C_t \leftarrow \max(C_t - 1,\ 1)$ \Comment{reclaim more aggressively}
\EndIf
\end{algorithmic}
\end{algorithm}

\subsection{Epoch Protocol and Optimistic Migration}
\label{sec:impl-epoch}

As shown in Figure~\ref{fig:thread-flow} the OC
coordinates migration through three states (\texttt{INACTIVE}, \texttt{PREPARE},
and \texttt{ACTIVE}) using a global epoch counter and a Thread Activity Index (TAI) - a per-thread slot that lets the OC verify, without blocking threads, that every active thread has observed the new epoch before migration begins.

\paragraph{Epoch Transitions.}
In the \texttt{INACTIVE} state, the ATC updates are disabled; dereferences only set access bits.
When the OC initiates migration, it increments the epoch counter and enters \texttt{PREPARE} state \circledBlue{1}. Threads entering public methods record the current epoch in the TAI (a small array indexed by thread-ID hash); exits clear their slot.
The OC repeatedly scans the TAI; once every non-empty slot reflects the new epoch \circledBlue{2}, all active threads have enabled ATC tracking, and the OC enters the \texttt{ACTIVE} state . After migration completes, the OC returns to \texttt{INACTIVE} state \circledBlue{3}. Crucially, threads never block; they simply record epoch participation, while the OC performs convergence checks.

\paragraph{Optimistic Data Migration.}
Within \texttt{ACTIVE} states, the OC migrates objects using an Optimistic Data Migration (ODM) protocol (Algorithm~\ref{odm-algo}), similar in spirit to Optimistic Concurrency Control (OCC), where a job is performed first and verified later, essentially acting as a weak transaction.
For a candidate with ATC=0 and migration-lock clear, the OC: (1) atomically sets the migration-lock bit on the guide slot; (2) allocates space in the target heap via SAMA and copies the object; (3) constructs a new guide (new address, heap ID, reset CIW, cleared lock) and publishes it with a single commit CAS, the same atomic both installs the new address and releases the lock. If the commit CAS fails, a concurrent dereference has changed the guide and the migration is abandoned (the OC explicitly clears the lock it set; see Algorithm~\ref{odm-algo}).

\begin{algorithm}[t]
\footnotesize
\caption{Optimistic Data Migration (ODM)}
\label{odm-algo}
\begin{algorithmic}[1]
\Procedure{MigrateObject}{$ptr$, $targetHeap$} \Comment{$ptr$: address of guide slot}
    \State $guide \leftarrow \textsc{AtomicLoad}(ptr)$ \Comment{Load object guide}
    \If{$\textsc{RefCount}(guide) > 0$}
        \State \Return $false$ \Comment{Object in use}
    \EndIf
    \State $srcAddr \leftarrow \textsc{ExtractAddr}(guide)$
    \State $srcHeap \leftarrow \textsc{ExtractHeapType}(guide)$ \Comment{Get current heap type}
    \State $\textsc{SetMigrationLock}(ptr)$ \Comment{CAS; mark as migrating}
    \State $size \leftarrow \textsc{GetSize}(srcAddr)$
    \State $dstAddr \leftarrow \textsc{samalloc}(size, targetHeap)$
    \If{$dstAddr = null$}
        \State $\textsc{ClearMigrationLock}(ptr)$
        \State \Return $false$ \Comment{Allocation failed}
    \EndIf
    \State $\textsc{Copy}(dstAddr, srcAddr, size)$ \Comment{Copy data}
    \State $newGuide \leftarrow \textsc{CreateGuide}(guide, dstAddr, targetHeap)$ \Comment{Inherits lock from guide metadata}
    \State $\textsc{ClearMigrationLock}(newGuide)$ \Comment{Local clear of inherited lock; commit via CAS below}
    \State $success \leftarrow \textsc{AtomicCAS}(ptr, guide, newGuide)$
    \If{$success$}
        \State $\textsc{Free}(srcAddr, srcHeap)$ \Comment{Release old memory}
        \State \Return $true$
    \Else
        \State $\textsc{Free}(dstAddr, targetHeap)$ \Comment{Rollback}
        \State $\textsc{ClearMigrationLock}(ptr)$ \Comment{CAS; release lock we set}
        \State \Return $false$ \Comment{Migration failed, guide changed}
    \EndIf
\EndProcedure
\end{algorithmic}
\end{algorithm}

\begin{table}[t]
\centering
\small
\setlength{\tabcolsep}{5pt}
\begin{tabular}{@{}clcccl@{}}
\toprule
\textbf{Time} & \textbf{Actor} & \textbf{Action} & \textbf{ATC} & \textbf{Lock} & \textbf{Outcome} \\
\midrule
\multicolumn{6}{l}{\textit{(a) Concurrent dereference aborts migration:}} \\
$t_0$  & OC     & read guide        & 0 & 0 & eligible \\
$t_1$  & OC     & CAS: set lock     & 0 & 1 & copying begins \\
\rowcolor{gray!10}
$t_2$  & Thread & dereference       & 1 & 0 & lock cleared \\
$t_3$  & OC     & CAS: commit       & \multicolumn{2}{c}{\textit{mismatch}} & \textbf{aborted} \\
\midrule
\multicolumn{6}{l}{\textit{(b) No interference, migration succeeds:}} \\
$t_0$  & OC     & read guide        & 0 & 0 & eligible \\
$t_1$  & OC     & CAS: set lock     & 0 & 1 & copying begins \\
$t_2$  & OC     & CAS: commit       & 0 & 0 & \textbf{success} \\
\bottomrule
\end{tabular}
\caption{\textbf{Race between migration and concurrent access.}
\small{(a) A dereference at $t_2$ modifies the guide, causing the OC's commit CAS to fail.
(b) When no thread intervenes, both CAS operations succeed and the object moves.}
}
\vspace{-3mm}
\label{tab:migration-race}
\end{table}

\paragraph{Race Resolution.}
Table~\ref{tab:migration-race} illustrates a race between migration and access.
Every dereference performs an atomic CAS that sets the access bit and \emph{clears}
the migration-lock bit. If a thread accesses object $x$ while the OC is copying it
($t_2$), the guide changes, the commit CAS fails ($t_3$), and $x$ remains at its
original address. The thread always sees valid data; concurrent access supersedes migration.

\subsection{Kernel Page Reclamation Optimization}
\label{sec:impl-kernel}

When objects migrate to the COLD heap and the Object Collector issues \texttt{MADV\_PAGEOUT}, Linux's page reclamation path becomes the bottleneck for efficient memory tiering.
The default path in \texttt{shrink\_folio\_list()} processes each page individually: it clears the page table entry (PTE), issues a TLB flush or shootdown that triggers an inter-processor interrupt (IPI) to every core, and submits the page to the block I/O layer.
This fine-grained approach creates substantial overhead when reclaiming large regions, as the cumulative cost of per-page TLB invalidations and IPIs degrades performance even for threads accessing unrelated hot data.

We modify \texttt{shrink\_folio\_list()} to batch these operations: it aggregates pages (up to a full PMD spanning 512 base pages), clearing each PTE and marking it for pageout, but defers TLB invalidation until a complete batch is prepared, then issues one flush over the entire range and submits all pages together for I/O.
By amortizing TLB shootdowns and I/O submissions, this reduces IPIs by more than 99\% when demoting 10\,GiB of memory versus the unmodified kernel.
Beyond local tiering, batched invalidation also addresses a scalability bottleneck in RDMA-based far-memory systems~\cite{aifm, leap,mira}, where per-page shootdowns cause IPI storms whose latencies grow super-linearly with thread count; aggregating across hundreds of pages lets the reclamation path scale for both CXL and RDMA backends.

\subsection{Compiler Passes for Guide Management}
\label{sec:llvm-pass}

\begin{figure}[t]
\centering
\begin{lstlisting}[language=C++,
  basicstyle=\ttfamily\scriptsize,
  keywordstyle=\bfseries,
  commentstyle=\color{gray},
  stringstyle=\color{black},
  emph={createTAG,addToTAG,destroyTAG},
  emphstyle={\color{initoutline}\bfseries},
  emph={[2]Guide},emphstyle={[2]\color{guideoutline}\bfseries},
  showstringspaces=false,
  numbers=left,numberstyle=\tiny\color{gray},numbersep=5pt,
  frame=single,rulecolor=\color{black!25},
  backgroundcolor=\color{black!4},
  xleftmargin=2.2em,xrightmargin=3pt,framexleftmargin=2.2em,framesep=4pt,
  aboveskip=2pt,belowskip=2pt,lineskip=-0.5pt]
void HashMap::get(int idx) {
    createTAG();                   // public entry
    Guide<char>& v = buckets[idx]; // HashMap member
    addToTAG(&v);                  // before deref
    if (*v > 37) { destroyTAG(); return; }
    addToTAG(&v);  *v = 42;
    addToTAG(&v);
    std::cout << "Value: " << *v << std::endl;
    destroyTAG();                  // public exit
}
\end{lstlisting}
\caption{\textbf{Compiler transformation.} \small{The developer marks one pointer field of \texttt{HashMap}; the pass rewrites its type from \texttt{char*} to \texttt{Guide<char>}, brackets the public method with \texttt{createTAG}/\texttt{destroyTAG}, and inserts \texttt{addToTAG} before each guide dereference. The \texttt{buckets} array is unmarked and stays in memory, so indexing itneeds no guard; only the guide dereference (\texttt{*v}) does. Callers invoke \texttt{HashMap::get} with a primitive key (no \texttt{Guide} crosses the public-method boundary), so caller code neither holds nor dereferences a \texttt{Guide}. ATC increments inside \texttt{addToTAG} only when the TAG state is not \texttt{INACTIVE}.}}
\label{fig:code-transformation}
\vspace{-3mm}
\end{figure}

\SYSTEM{} uses three complementary compiler passes, implemented on Clang and LLVM, to convert raw pointers to guides and insert the TAG/ATC instrumentation, confining developer effort to marking which pointer fields are managed.
\myparagraph{Visibility extraction} A Clang frontend pass ensures TAGs are created only at public boundaries by using a \texttt{RecursiveASTVisitor} to find public methods that serve as data-structure entry points (e.g., \texttt{get}, \texttt{set}, \texttt{delete}) and records their visibility for later stages.
\myparagraph{Pointer-to-guide conversion} A second Clang pass uses the rewriter to convert the developer-listed pointer declarations and usages into guides, giving precise control over which objects \SYSTEM{} manages.
\myparagraph{IR instrumentation} The third pass runs on the LLVM IR. 
 A fixed-point analysis first identifies the set of functions that directly use guides (via conversions, destructor calls, operator overloads, or assignments), propagates this set through the call graph to find functions that touch guides indirectly, and combines it with the visibility data from Pass 1 to classify each function. It then inserts \texttt{createTAG}/\texttt{destroyTAG} at the entry and exit of public functions that touch guides, and \texttt{addToTAG} before each guide dereference in all transformed functions (public and private). This restricts TAG creation to public boundaries while still tracking every access throughout the call stack (Figure~\ref{fig:code-transformation}).
\section{Evaluation}
\label{sec:evaluation}

We evaluate \SYSTEM{} along four dimensions:

\myparagraph{E1: Effectiveness} Does \SYSTEM{} reduce hotness fragmentation and memory footprint across different data structures and workloads? (§\ref{sec:eval-effectiveness})

\myparagraph{E2: Backend synergy} How much does \SYSTEM{} improve the effectiveness of existing page-based reclamation and tiering backends? (§\ref{sec:eval-backends})

\myparagraph{E3: Overhead and scalability} What runtime overheads do tracking and migration introduce, and how do they scale with thread count? (§\ref{sec:eval-overhead})

\myparagraph{E4: Dynamic behavior} How does \SYSTEM{} adapt to changing hotsets in long-running, real-world workloads? (§\ref{sec:eval-traces})

\subsection{Experimental Setup}
\label{sec:eval-setup}

All experiments run on an Intel Xeon Gold 5218 (16 cores, SMT disabled) server configured in \texttt{performance} governor mode with Ubuntu~22.04 and Linux kernel~6.12.
The memory subsystem comprises two tiers: a fast tier of 2$\times$16\,GB DDR4 DRAM modules at 2400\,MHz, and a slow tier of 4$\times$128\,GB Intel Optane DC Persistent Memory 100 modules at 2666\,MHz.
All six memory devices occupy distinct channels to avoid interface.
We configure Optane PMEM in \texttt{Memory Mode} and expose it as a separate NUMA node, creating a two-tier hierarchy representative of emerging CXL-attached memory deployments~\cite{pond, tpp}.
The Optane tier provides approximately 2.5$\times$ higher access latency than local DRAM~\cite{emperical-guide-optane}, consistent with first-generation CXL memory characteristics~\cite{demystifying-cxl}.
For reclamation experiments requiring swap, we use a 512\,GB Intel P4800X Optane SSD.

\begin{table}[t]
\centering
\scriptsize
\begin{tabular}{lll}
\toprule
\textbf{Structure} & \textbf{Concurrency Control} & \textbf{Used In} \\
\midrule
HashTable Harris \cite{harris} & Lock-free algorithm & NGINX \\
HashTable Pugh \cite{pugh} & Fine-grained r/w lock & Redis, Memcached \\
HashTable Java CHM \cite{chm} & Segmented bucket locks & Linux kernel, HAProxy \\
\midrule
SkipList Coarse & Global lock & LevelDB/RocksDB \\
SkipList Fraser \cite{fraser} & Lock-free algorithm & Redis Sorted Sets \\
SkipList Herlihy \cite{herlihy} & Optimistic fine-grained & Cassandra, CockroachDB \\
\midrule
B+Tree Coarse & Global lock & SAP HANA \\
B+Tree OCC & OCC w/ epoch reclaim & VoltDB index \\
MassTree \cite{masstree} & OCC + RCU & MICA, Silo \\
\midrule
Adaptive Radix Tree \cite{art} & Fine-grained r/w lock & DuckDB, PostgreSQL \\
\bottomrule
\end{tabular}
\caption{\textbf{Concurrent data structures evaluated with \SYSTEM{}.} 
    \small{These structures span lock-free, fine-grained, and coarse-grained concurrency mechanisms, demonstrating \SYSTEM{}'s compatibility with diverse synchronization approaches.}
}
\vspace{-3mm}
\label{tab:ds}
\end{table}

\myparagraph{CrestDB testbed}
We implemented CrestDB, a concurrent in-memory key-value store, to evaluate \SYSTEM{} across diverse data structures and concurrency mechanisms.
CrestDB integrates ten high-performance data structures spanning the concurrency-control spectrum (Table~\ref{tab:ds}), from lock-free algorithms to global locks.
Many structures are borrowed from ASCYLIB~\cite{ascylib}, which provides production grade implementations.
This diversity demonstrates that \SYSTEM{}'s compiler instrumentation and migration protocol are compatible with a wide range of synchronization approaches without structure-specific modifications.
All data structures maintain guide pointers to key and value objects; CrestDB deep-copies inserted data, ensuring \SYSTEM{} manages the authoritative copy rather than application-held aliases.
Unless otherwise noted, CrestDB runs with six server threads and six client threads.

\myparagraph{Workloads}
We use the YCSB benchmark suite with Zipfian-distributed keys to model skewed access patterns typical of production workloads.
To evaluate real-world adaptivity, we replay production traces from Meta (CacheLib~\cite{cachelib-osdi}, DBbench~\cite{benchmark-rocksdb-fb}) and Twitter (Cluster~7, Cluster~23)~\cite{twitter-traces}.

\myparagraph{Controller parameters}
Unless otherwise noted, the Object Collector scans every 120\,s, targets a 1\% promotion rate, and adjusts $C_t$ by $\pm1$ window per scan within $1\le C_t\le32$ (§\ref{sec:address-space-org}); the conservative 1\% target follows production compressed-memory deployments~\cite{TMO, zswap}.
The optimal rate is hardware-dependent (faster tiers tolerate higher cold-access fractions), and hardware-aware tuning is future work. §\ref{sec:eval-traces} examines $C_t$ dynamics over time.

\subsection{Effective Address-Space-Engineering (E1)}
\label{sec:eval-effectiveness}

We first evaluate whether \SYSTEM{} achieves its core objective: reducing hotness fragmentation and converting unreclaimable cold data into reclaimable pages.
We run CrestDB with all ten data structures under YCSB workloads A, B, and C with Zipfian keys.
We load 10M keys with 30-byte keys and 1024-byte values, creating a 13\,GiB dataset.
Unless stated otherwise, all results in this section are measured after \SYSTEM{} has converged (promotion rate below the 1\% target).

\begin{figure}[t]
    \centering

    \includegraphics[width=0.47\textwidth]{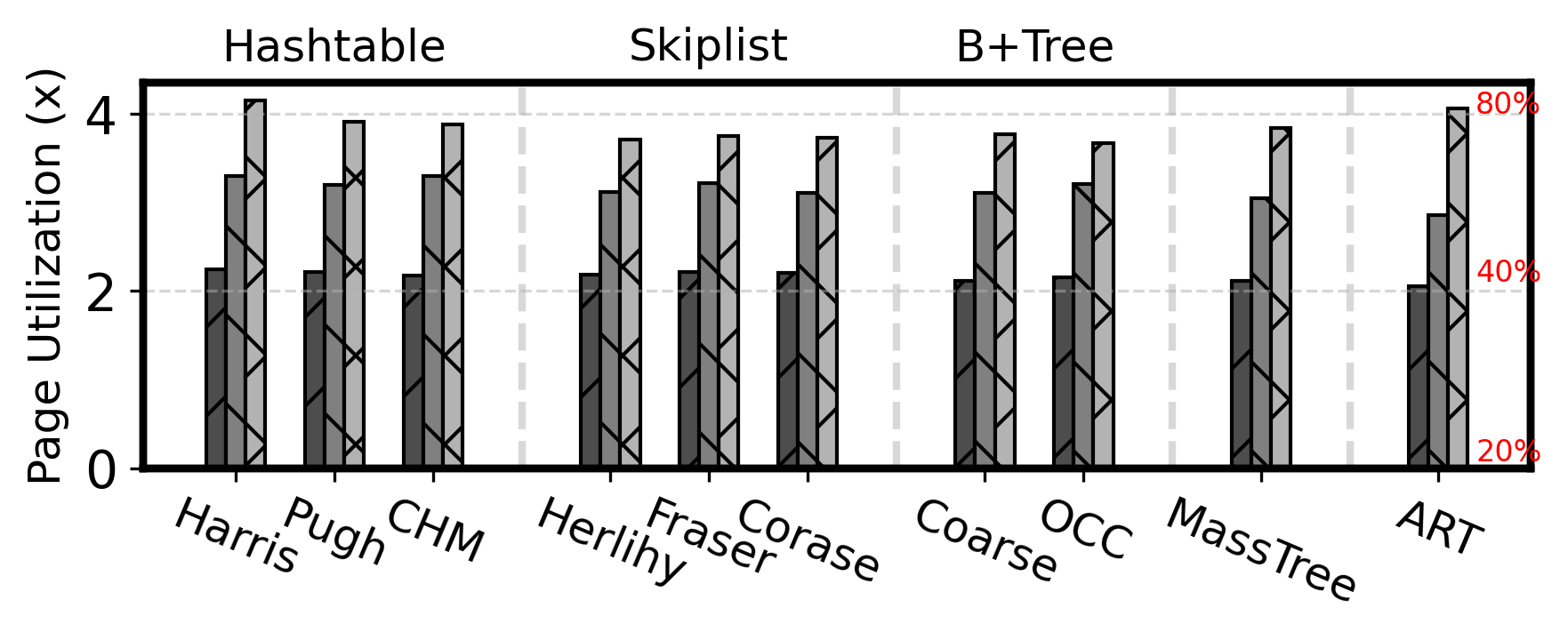}
    \vspace{-0.12em}
    \includegraphics[width=0.47\textwidth]{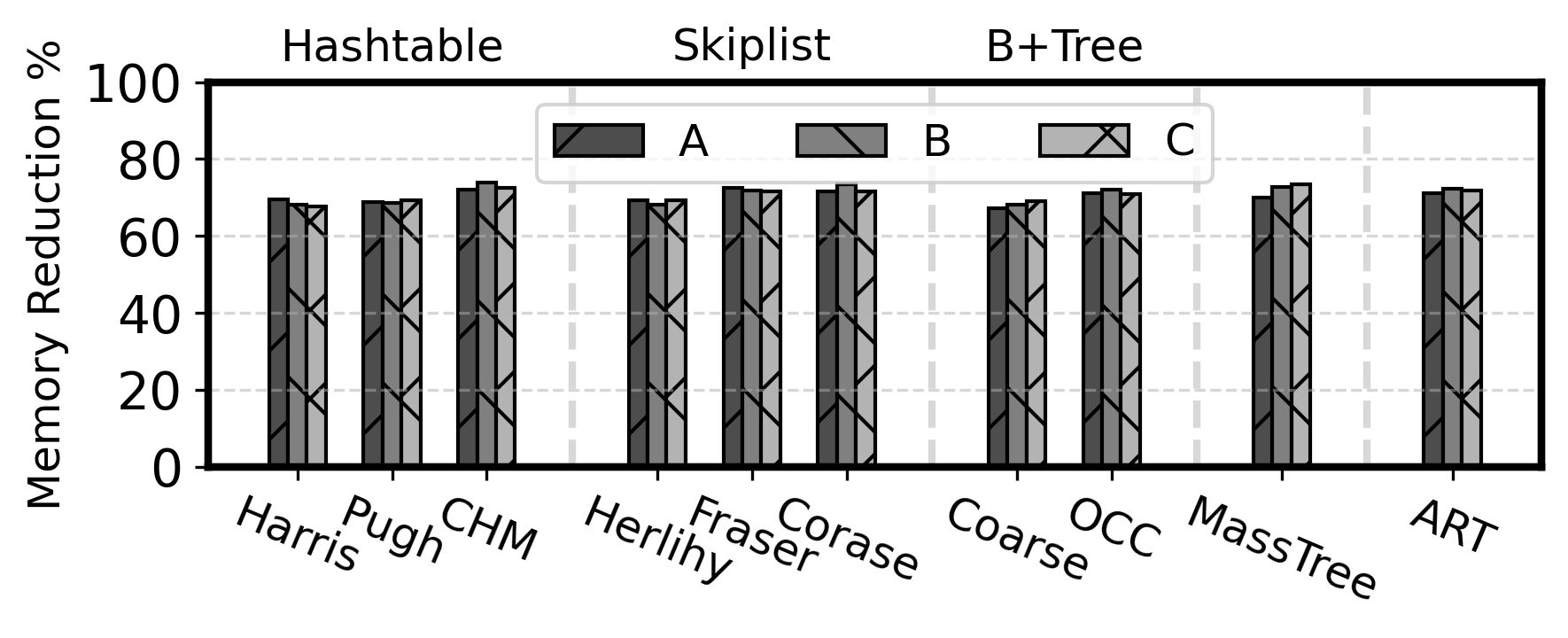}
    \vspace{-0.2em}
    \caption{\textbf{\SYSTEM{} effectiveness (YCSB, 10M keys).}
    \small{
        Top: Page utilization improvement relative to the baseline allocator; \SYSTEM{} increases utilization by 2--4$\times$ across workloads and data structures.
        Bottom: RSS reduction after \SYSTEM{} pages out the COLD heap (via Kswapd). Memory footprint shrinks by 65--72\%.}}
    \label{fig:effectiveness}
    \vspace{-4mm}
\end{figure}

\myparagraph{Page utilization}
Figure~\ref{fig:effectiveness}(a) reports page utilization before and after \SYSTEM{} reorganizes objects into NEW, HOT, and COLD heaps.
Initially, the data structures exhibit 18--20\% utilization when measured over 120\,s windows: most pages contain only a handful of accessed cache lines, reflecting the hotness fragmentation observed in production traces (§\ref{sec:motivation}).

After three scan intervals, the Object Collector classifies objects based on guide access bits and migrates them into HOT or COLD heaps.
Compared to the baseline, \SYSTEM{} improves page utilization by 2$\times$ for workload~A (50\% writes), approximately 3$\times$ for workload~B (5\% writes), and up to 4$\times$ for the read-only workload~C.
Across data structures, absolute utilization after convergence ranges from roughly 40\% (workload~A) to 80\% (workload~C).

The variation across workloads reflects how the NEW and HOT heaps interact.
Workload~C has no updates: once objects are classified as hot, they remain in the HOT heap and no new objects are allocated there.
Nearly all accessed bytes end up densely packed in a small number of HOT pages, and utilization approaches 80\%.
In workload~B, occasional updates allocate fresh values in NEW, so the working set splits between NEW and HOT and overall utilization stabilizes around 60--70\%.
Workload~A performs frequent updates, continuously injecting new objects into NEW.
Utilization still roughly doubles, but cannot reach read-only levels because a larger fraction of hot objects reside in NEW during their initial epochs.

The consistency of the improvement across ten structurally different data structures shows that \SYSTEM{}'s benefits derive from object-temperature clustering rather than data-structure-specific layout optimizations.

\myparagraph{Memory footprint}
Higher page utilization translates directly into reclaimable cold memory.
Once the promotion rate falls below the 1\% target---typically after 3--4 scan intervals (6--8 minutes) for YCSB---\SYSTEM{} proactively issues \texttt{madvise(MADV\_PAGEOUT)} on the COLD heap.
Figure~\ref{fig:effectiveness}(b) shows the resulting RSS reduction relative to a baseline without reclamation.

Across all data structures and workloads, \SYSTEM{} reduces RSS by 65--72\%.
For workload~B with 10M keys, the baseline uses 12.4\,GiB; after \SYSTEM{} converges and pages out COLD, RSS drops to 3.5--4.0\,GiB.
Because COLD pages contain almost exclusively inactive objects, proactive paging does not cause swap-in storms or noticeable throughput degradation (we quantify overheads in §\ref{sec:eval-overhead}).

\textit{\textbf{Takeaway \#1:} \SYSTEM{} improves page utilization across all data structures as it tracks object hotness without the semantic knowledge of each structure. This enables uniform hotness fragmentation reduction across diverse concurrency mechanisms.}

\subsection{Backend Synergy (E2)}
\label{sec:eval-backends}

The memory savings demonstrated in §\ref{sec:eval-effectiveness} are valuable only if OS tiering backends can exploit them without degrading performance.
We now show that \SYSTEM{} enables existing reclamation and tiering systems to achieve aggressive memory savings while preserving throughput.

\subsubsection{Paging-based Reclamation Backends}
\label{sec:eval-reclaim}

We run CrestDB with Mass\-Tree under YCSB-C in a memory-constrained configuration: the workload has a 13\,GiB footprint but an active working set of approximately 4\,GiB.
We compare four reclamation strategies, each with and without \SYSTEM{} as a frontend:

\begin{itemize}[leftmargin=*, topsep=2pt, itemsep=1pt]
    \item \textbf{Kswapd:} Linux's background reclaimer, triggered by memory pressure from a co-located process.
    \item \textbf{Cgroup:} Memory limit set to working-set size (4\,GiB), forcing aggressive reclamation.
    \item \textbf{TMO:} Meta's PSI-based proactive reclaimer~\cite{TMO}.
    \item \textbf{\SYSTEM{} Hinted:} Proactive \texttt{MADV\_PAGEOUT} on the COLD heap after convergence.
\end{itemize}

\noindent
Throughout the evaluation, ~\SYSTEM{} denotes the frontend alone, address-space reorganization with proactive paging disabled, so combinations like ~\SYSTEM{}+TMO isolate the benefit of reorganization from any hinting. ~\SYSTEM{} Hinted additionally issues \texttt{MADV\_PAGEOUT} on the COLD heap.

\begin{figure}[t]
    \centering
    \includegraphics[width=0.47\textwidth]{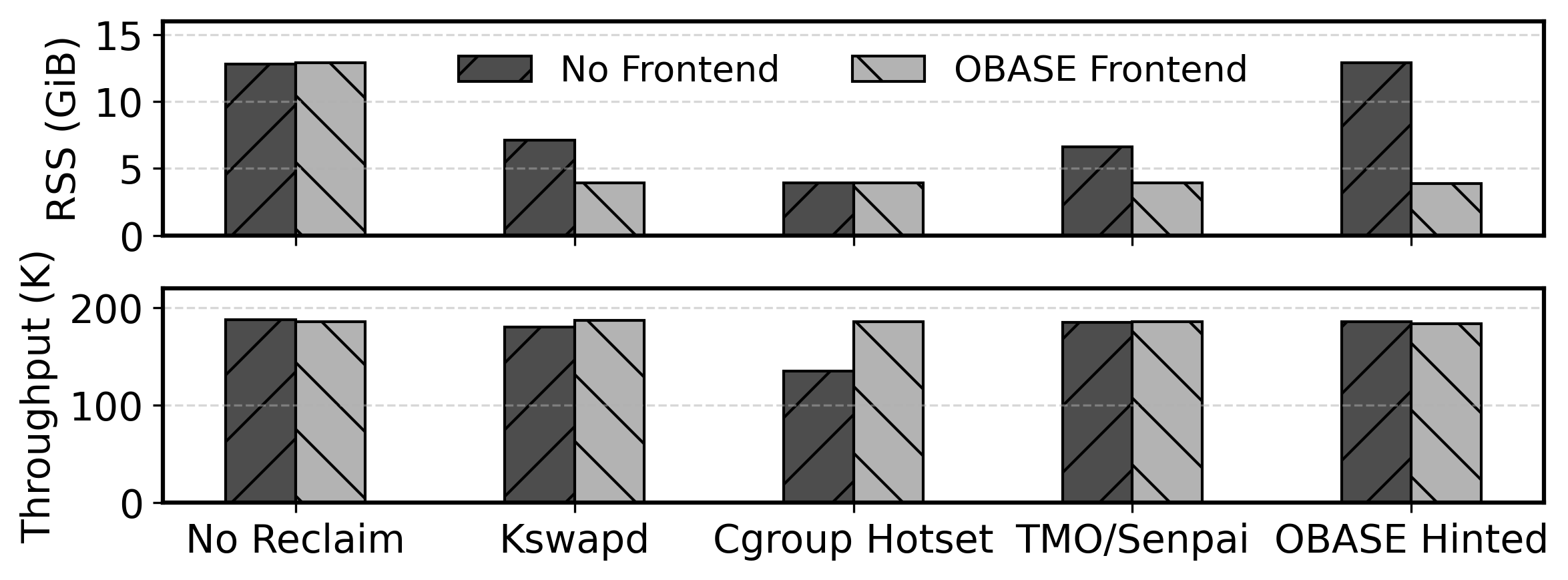}
    \vspace{-2mm}
    \caption{\textbf{\SYSTEM{} with reclamation backends (YCSB-C, MassTree).}
    \small{Top: RSS after convergence. Bottom: throughput.
    Without \SYSTEM{}, backends face a trade-off between memory savings and performance.
    With \SYSTEM{} (hatched bars), all backends achieve near-optimal memory savings with minimal throughput loss.}}
    \label{fig:reclaim-backend}
    \vspace{-4mm}
\end{figure}

Figure~\ref{fig:reclaim-backend} reveals a fundamental trade-off that \SYSTEM{} resolves.
\textbf{Without \SYSTEM{}}, backends must choose between memory efficiency and performance.
Kswapd reduces RSS from 13\,GiB to 7\,GiB (1.8$\times$) with no throughput loss, but leaves 3\,GiB of cold data trapped in mixed-temperature pages due to its page level view through PTE scans.
Cgroup reaches 4\,GiB (3.2$\times$), but throughput collapses by 38\% as the kernel inevitably evicts hot objects.
TMO achieves a 6.5\,GiB RSS (2$\times$) with no throughput loss, but cannot reclaim further because PSI signals page-level pressure regardless of object-level coldness.

\textbf{With \SYSTEM{}}, all backends reach 4\,GiB RSS---matching the most aggressive policy---with no throughput degradation.
Kswapd now reclaims COLD pages preferentially. Cgroup no longer thrashes because evicted pages contain genuinely cold objects. TMO's PSI probes no longer encounter mixed-temperature pages that resist reclamation.
\SYSTEM{} Hinted achieves the same result proactively, without relying on any backend policy.


\textit{\textbf{Takeaway \#2:} \SYSTEM{} achieves aggressive reclaimation while preserving application performance.}

\subsubsection{Tiering Backends}
\label{sec:eval-tiering}
To demonstrate the memory tiering benefits of reduced hotness fragmentation, we evaluate \SYSTEM{} with three page based migration systems: TPP~\cite{tpp}, AutoNUMA~\cite{autonuma}, and Memtis~\cite{memtis}.

\myparagraph{Setup}
We load CrestDB (MassTree) with 50M keys (30-byte keys, 1024-byte values), a 67\,GiB dataset.
We vary the DRAM:CXL ratio across three configurations: 1:4 (14.8\,GiB DRAM), 1:8 (7.4\,GiB DRAM), and 1:16 (3.9\,GiB DRAM), with the remaining capacity on Optane PMEM.
As in ~\cite{memtis}, performance is normalized to a baseline where all data resides on the slow tier; values above 1.0$\times$ indicate speedup from effective DRAM use.

\myparagraph{The hot-set mismatch}
Without \SYSTEM{}, the working set spans 16.3\,GiB of pages at 21\% utilization--- slightly larger than the 1:4 DRAM budget of 14.8\,GiB.
No configuration can keep all accessed data in DRAM.
With \SYSTEM{}, the same logical working set compacts to 6.33\,GiB at 57\% utilization, fitting comfortably at 1:4 and 1:8 ratios.

\begin{figure}[t]
    \centering
    \includegraphics[width=0.47\textwidth]{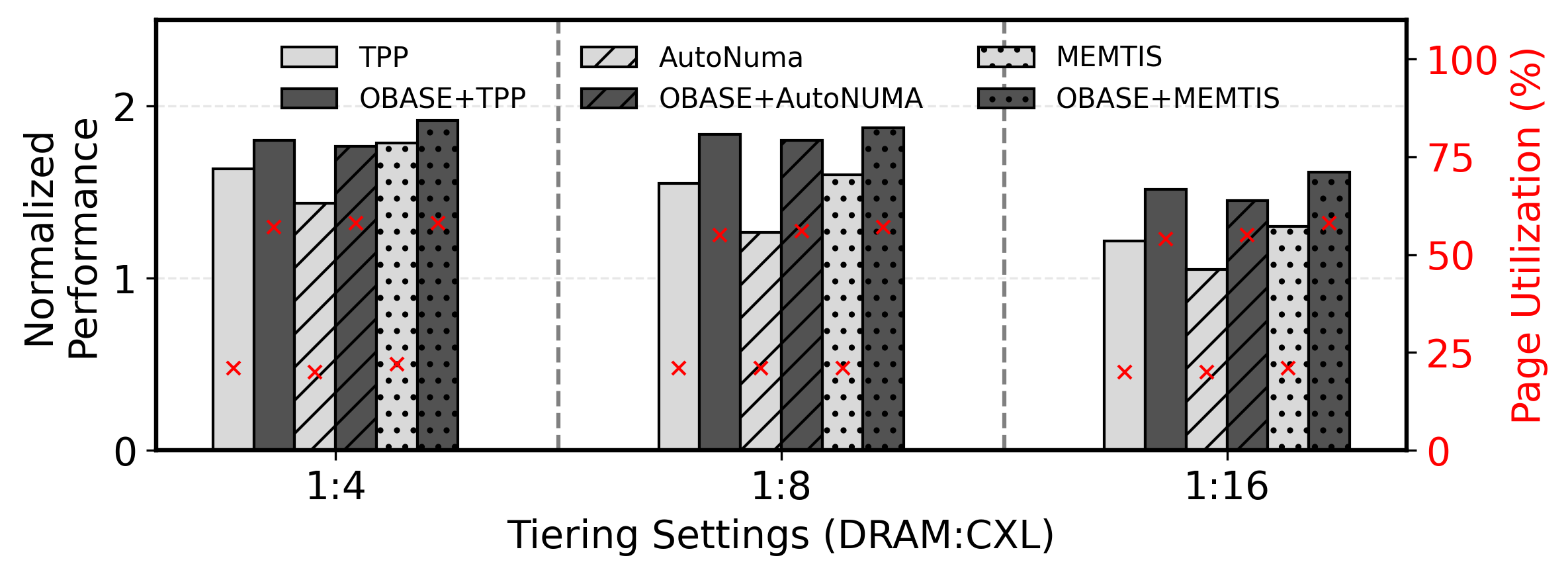}
    \vspace{-2mm}
    \caption{\textbf{\SYSTEM{} with tiering backends (YCSB-B, MassTree, 50M keys).}
    \small{Throughput normalized to CXL-only baseline (higher is better).
    Without \SYSTEM{}, performance degrades as DRAM shrinks because the hot set cannot fit.
    With \SYSTEM{}, the hot set compacts, enabling stable performance even at 1:16.}}
    \label{fig:tiering-backend}
    \vspace{-4mm}
\end{figure}

\myparagraph{TPP}
TPP uses hysteresis-based promotion and proactive demotion to manage DRAM headroom.
Figure~\ref{fig:tiering-backend} shows TPP achieves 1.65$\times$ speedup at 1:4, degrading to 1.25$\times$ at 1:16 as the DRAM budget shrinks below the fragmented hot set.
With \SYSTEM{}, TPP reaches 1.85$\times$ at 1:4 and retains 1.45$\times$ at 1:16---a 16\% improvement at the most constrained ratio.

\myparagraph{AutoNUMA}
AutoNUMA promotes any accessed remote page immediately, without hysteresis, causing thrashing when the hot set exceeds DRAM capacity.
Without \SYSTEM{}, AutoNUMA underperforms TPP by 15--20\%, achieving only 1.05$\times$ at 1:16---barely better than CXL-only.
With \SYSTEM{}, AutoNUMA matches TPP-alone performance: at 1:8, \SYSTEM{}+AutoNUMA (1.6$\times$) exceeds TPP alone (1.55$\times$).
When pages are uniformly hot or cold, even naive promotion decisions become correct.

\myparagraph{Memtis}
Memtis uses hardware sampling (PEBS) to identify hot pages, achieving the best baseline results: 1.8$\times$ at 1:4 and 1.55$\times$ at 1:16.
With \SYSTEM{}, Memtis improves to 1.95$\times$ and 1.7$\times$, respectively.
The gains are smaller (3--10\% vs.\ 12--29\% for TPP/AutoNUMA) because Memtis already captures much of the page-level signal---but even the most sophisticated page-level policy benefits from object-level reorganization.

Because \SYSTEM{} reduces the effective hot set by 2.5$\times$, ratio \textbf{1:X} performs comparably to baseline at \textbf{1:(X/2)}. For example, \SYSTEM{} + TPP at 1:16 reaches 1.45$\times$, within $\sim$6\% of TPP alone at 1:8 (1.55$\times$). For operators, this means the same performance with half the DRAM, or double the effective capacity of existing tiered deployments.

\textit{\textbf{Takeaway \#3:} \SYSTEM{} compacts the hot set so it fits in smaller DRAM budgets, making page-based tiering backends effective.}

\subsection{Overhead and Scalability (E3)}
\label{sec:eval-overhead}

Given the memory savings and backend improvements demonstrated above, we now quantify \SYSTEM{}'s runtime costs.

\subsubsection{Steady-State Overhead}

\begin{figure}[t]
    \centering
    \includegraphics[width=0.47\textwidth]{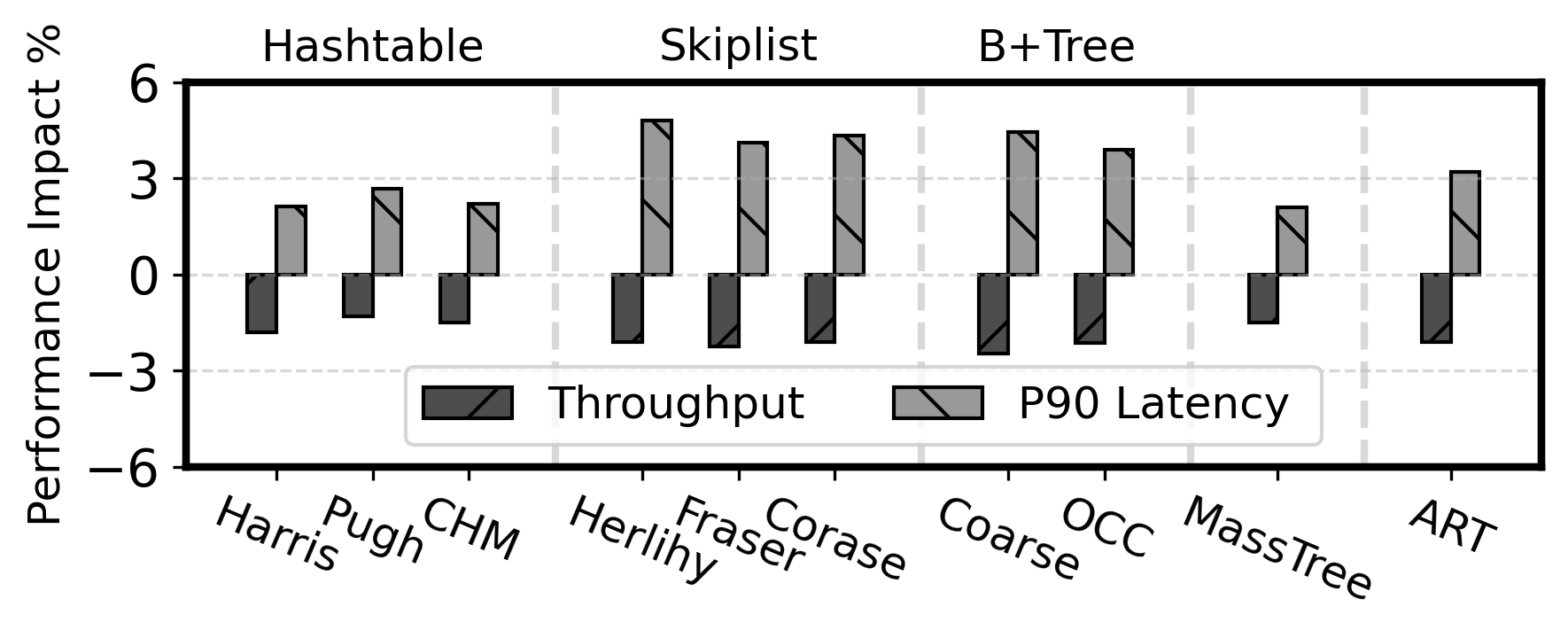}
    \vspace{-0.2em}
    \includegraphics[width=0.47\textwidth]{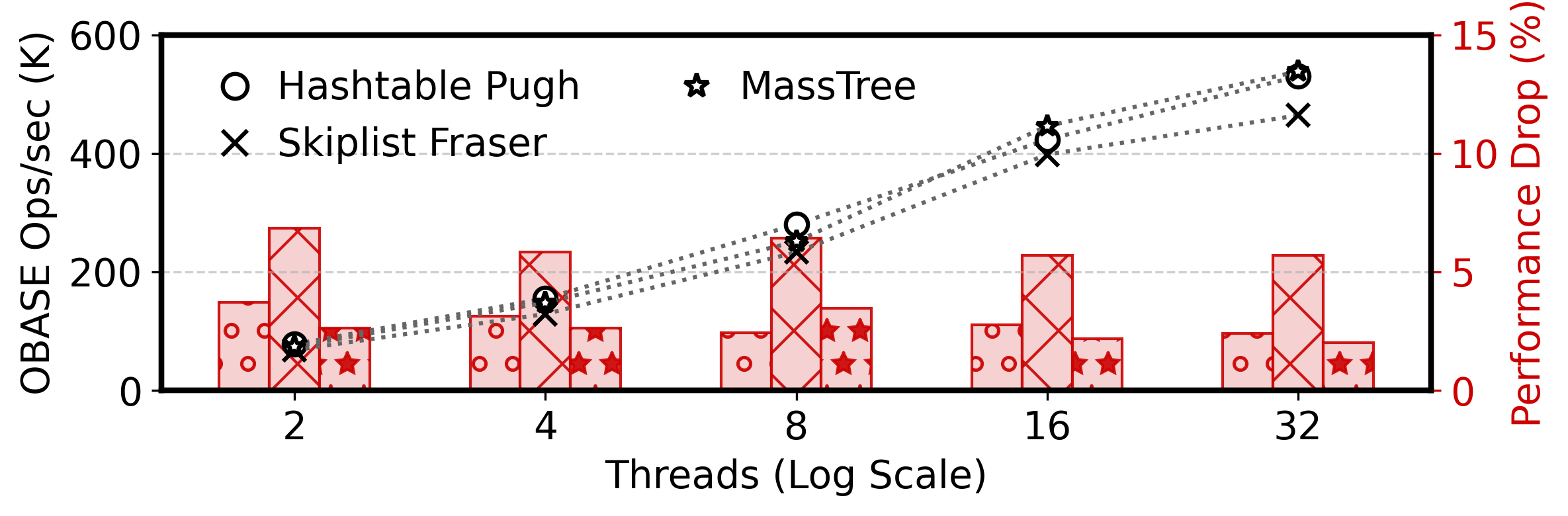}
    \vspace{-0.2em}
    \caption{\textbf{\SYSTEM{} overhead and scalability (YCSB, 10M keys).}
    \small{
        Top: Throughput and p90 latency overhead across data structures. Overhead ranges from 1.5--5\% depending on structure.
        Bottom: Scalability from 2 to 32 threads. Bars show absolute throughput; markers show overhead relative to baseline. Overhead remains bounded at 1--8\% with no upward trend.}}
    \label{fig:overhead}
    \vspace{-3mm}
\end{figure}

Figure~\ref{fig:overhead}(a) reports throughput and p90 latency overhead when no reclamation or tiering backend is active, normalized to an uninstrumented baseline.

On average, \SYSTEM{} reduces throughput by 2.5\% and increases p90 latency by 5\%.
Hash tables see the smallest impact (1.5--3\% throughput drop), while skiplists, B+Trees, and ART experience 3--5\% overhead.
This variation correlates with the number of guides touched per operation: hash table lookups dereference few nodes, whereas tree traversals visit more nodes and incur proportionally more tracking overhead.

The overhead has two main sources:
(1) a tagged-pointer read-modify-write on each guide dereference (4--5\,ns, comparable to an L1 cache hit), and
(2) TAG/ATC bookkeeping during \texttt{ACTIVE} migration epochs.
The Object Collector runs in a dedicated thread and consumes less than 1\% of CPU time.

\subsubsection{Thread Scalability}

A natural concern is whether \SYSTEM{}'s atomic operations become contention bottlenecks at higher thread counts.
We measure scalability by varying CrestDB server threads from 2 to 32 on three representative data structures spanning different synchronization mechanisms: Hashtable Pugh (fine-grained locking), Skiplist Fraser (lock-free), and MassTree (OCC with epoch reclamation).

Figure~\ref{fig:overhead}(b) shows that throughput scales and overhead remains bounded at 1--8\% regardless of thread count.
Critically, overhead does not increase with concurrency: all three data structures exhibit similar overheads at 2 and 32 threads.

This scalability follows from \SYSTEM{}'s design: guide metadata updates target per-object state (no cross-thread contention), the test-and-set optimization skips redundant writes for hot objects, TAGs are thread-local, and ATC increments occur only during brief \texttt{ACTIVE} epochs.
The 2--5\% overhead is measured against a DRAM-only baseline with no memory pressure—an idealized scenario that production systems rarely enjoy.
In the tiered-memory environments where \SYSTEM{} is designed to operate, the comparison reverses: as \S\ref{sec:eval-backends} showed, backends \emph{without} \SYSTEM{} suffer 10--38\% throughput loss from poor page-selection decisions, while backends \emph{with} \SYSTEM{} match DRAM-only performance.

\textit{\textbf{Takeaway \#4:} 
\SYSTEM{} imposes 1.5–5\% overhead across data structures and stays within 8\% from 2 to 32 threads, a modest cost relative to the backend improvements in §\ref{sec:eval-effectiveness}–\ref{sec:eval-backends}.
}

\subsection{Real World Traces}
\label{sec:eval-traces}

Synthetic workloads exercise controlled forms of skew, but production systems exhibit substantially more complex behavior: shifting hotsets, mixed read/write/delete ratios, and locality patterns that evolve over hours.
We therefore evaluate \SYSTEM{} on four real-world traces to assess whether its fragmentation-reduction benefits generalize beyond YCSB and whether the feedback controller adapts stably to long-term changes in access patterns.


We evaluate four traces that span a range of access patterns:
\begin{itemize}[leftmargin=*, topsep=2pt, itemsep=1pt]
    \item \textbf{Meta CacheLib}~\cite{cachelib-osdi}: Read-heavy (83\% GET) with gradually shifting popular keys.
    \item \textbf{DBench Mixgraph}~\cite{benchmark-rocksdb-fb}: Models Meta's ZippyDB with key-range locality across 30 prefixes. Read-heavy (85\% GET, 14\% PUT, 1\% SEEK).
    \item \textbf{Twitter Cluster~7}~\cite{twitter-traces}: High skew ($\alpha{=}1.07$) with small, concentrated working set of reads and writes.
    \item \textbf{Twitter Cluster~23}~\cite{twitter-traces}: Write-heavy (31\% SET, 30\% INCR) with low skew ($\alpha{=}0.274$) and deletes.
\end{itemize}

\noindent
These traces cover the spectrum from highly skewed (Cluster~7) to nearly uniform (Cluster~23), and from read-dominated (CacheLib) to write-heavy (Cluster~23).
We replay each trace on CrestDB with ART and measure memory reduction and page utilization improvement.

\begin{figure}[t]
    \centering
    \includegraphics[width=0.44\textwidth]{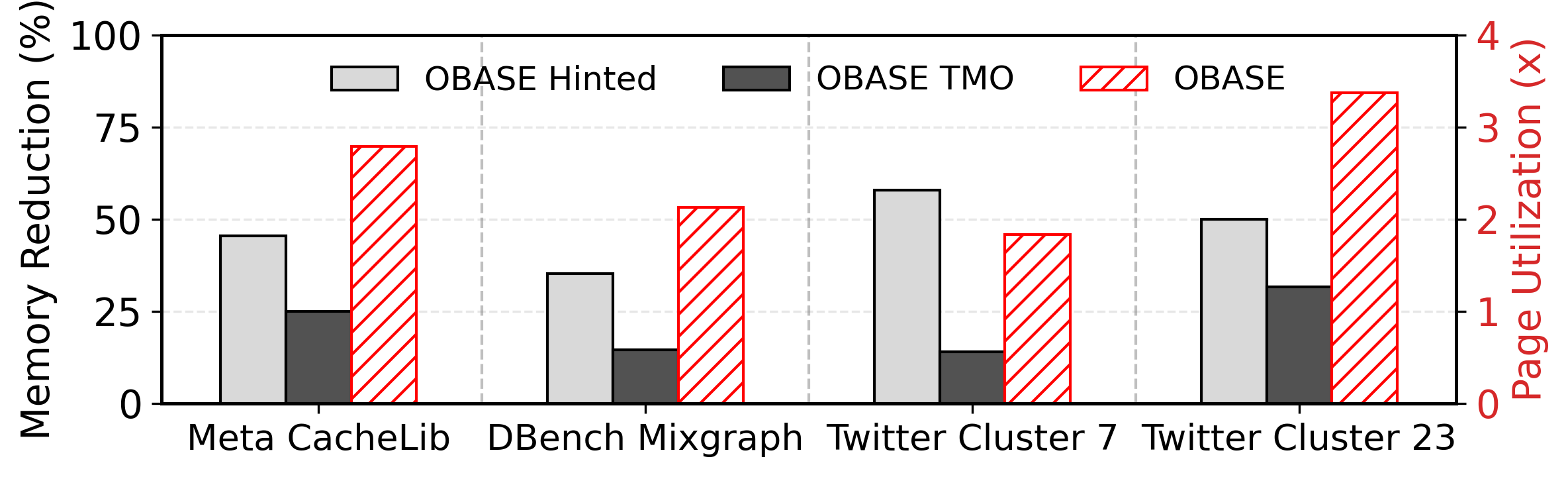}
    \vspace{-2mm}
    \caption{\textbf{Production trace results.}
    \small{Memory reduction (left axis): \SYSTEM{} Hinted vs.\ no-reclaim, and \SYSTEM{}+TMO vs.\ TMO-alone.
    Page utilization improvement (right axis, hatched) from address-space reorganization.}}
    \label{fig:trace-summary}
    \vspace{-4mm}
\end{figure}

\myparagraph{Page utilization (E1)}
Page utilization improves by 1.8--3.4$\times$ across traces.
Cluster~23 shows the highest gain (3.4$\times$) because its low skew disperses accesses across many keys, yielding very low baseline utilization.
Cluster~7's high skew naturally concentrates accesses, so the baseline is already reasonable and the relative improvement is smaller (1.8$\times$).

\myparagraph{Memory reduction (E1, E2 \& E4)}
Figure~\ref{fig:trace-summary} shows that \SYSTEM{} Hinted reduces RSS by 36--58\% compared to no reclamation.
Cluster~7 achieves the largest reduction (58\%) because its high skew concentrates the working set into fewer hot objects; DBench shows the smallest (36\%) because key-range locality spreads accesses more uniformly. 
Adding \SYSTEM{} to TMO provides 15--30\% additional savings relative to TMO-alone, demonstrating
that \SYSTEM{} and TMO are complementary: TMO identifies reclaimable pages, while \SYSTEM{} ensures those pages contain uniformly cold data.

\begin{figure}[t]
    \centering
    \includegraphics[width=0.47\textwidth]{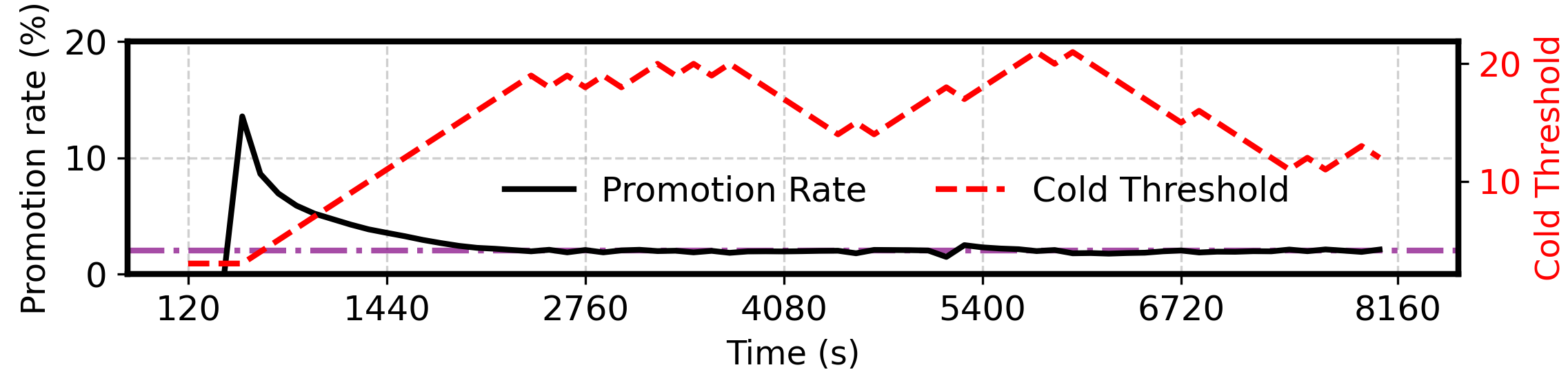}
    \vspace{-2mm}
    \caption{\textbf{Cold threshold adaptation (Meta CacheLib).}
    \small{Promotion rate (black) and cold threshold $C_t$ (red) over time. The controller automatically adjusts $C_t$ to maintain the 1\% target (purple).}}
    \label{fig:meta-adaptive}
    \vspace{-4mm}
\end{figure}

\myparagraph{Adaptive cold threshold (E4)}
Figure~\ref{fig:meta-adaptive} demonstrates \SYSTEM{}'s ability to adapt to workload 
dynamics over 2.3 hours of the Meta CacheLib trace.
At startup, the initial $C_t{=}3$ causes premature demotions, spiking the promotion 
rate to 14\%.
The controller increments $C_t$ each scan interval; within 25 minutes, $C_t$ reaches
18 and the promotion rate drops below the 1\% target. A high promotion rate during convergence \textit{does not indicate degraded performance}: COLD-heap pages remain in DRAM until a backend explicitly pages them out, and the promotion rate reflects workload behavior rather than backend decisions (\S\ref{sec:address-space-org}).

Around $t{=}5400$\,s, a workload shift causes a brief spike, as previously cold keys 
become active; the controller raises $C_t$ and the system re-converges as the new 
hotset stabilizes.
Throughout the trace, $C_t$ varies between 10--20 windows while maintaining the 
promotion rate near the target, demonstrating that \SYSTEM{} tracks workload evolution.

\textit{\textbf{Takeaway \#5:} \SYSTEM{} delivers consistent memory savings across real world workloads with diverse characteristics and dynamically adapts to shifting access patterns.}
\section{Related Work}

\myparagraph{Object-Level Management}
AIFM~\cite{aifm} and MIRA~\cite{mira} manage memory at object granularity for disaggregated far memory, bypassing OS page management for specialized user-space runtimes; AIFM builds on Shenango's green threads and kernel-bypass networking, with explicit \texttt{DerefScope} guards around far-memory accesses.
\SYSTEM{} is complementary: it acts as a \emph{frontend} that reorganizes the address space for unmodified OS \emph{backends}, optimizing \emph{what} to tier rather than the latency of remote access, so we do not compare directly.
The \texttt{DerefScope} contrast is instructive: it guards each individual far-memory dereference so the runtime can fault the object in, whereas \SYSTEM{}'s TAG/ATC scopes bracket a whole public operation as a quiescence check that permits migration only when no thread is active, closer to epoch-based reclamation than to per-access load barriers.

Other systems indirect through pointers for different ends.
Midas~\cite{midas} harvests idle memory for application-managed \emph{soft state}, dropping recomputable data under pressure, while Atlas~\cite{atlas-osdi} accelerates far-memory applications with a hybrid paging/runtime data plane; both optimize access to \emph{remote} memory, whereas \SYSTEM{} reorganizes \emph{local} resident objects by observed access intensity so page-based backends see uniformly hot or cold pages.
Alaska~\cite{handle} uses handle-based indirection for heap compaction but does not track object hotness or organize memory by access intensity.
ObjecTier~\cite{objectier} proposes a non-invasive object-consolidation framework with the same high-level vision as \SYSTEM{}, motivated by simulations and they defer implementation to future work.

\myparagraph{Allocation-Time Placement}
Static placement approaches \cite{xmem, atmem, data-spatial-locality, tcmalloc, llvm-pgho} make decisions at allocation time based on hints or profiling. However, object hotness is neither knowable at allocation nor stable over time(\S\ref{sec:motivation}). \SYSTEM{} tracks access patterns at runtime, enabling runtime migration. 

\myparagraph{Page-Level Tiering} TPP~\cite{tpp}, Memtis~\cite{memtis}, HawkEye~\cite{hawkeye}, and TMO~\cite{TMO} all operate at page granularity. Even efforts to shrink the tiering granularity (e.g., Memtis's dynamic 4KB classification~\cite{memtis}) are bounded by the page floor, since sub-page objects of differing hotness still share a 4KB page (\S\ref{sec:motivation}). \SYSTEM{} works below this floor by reorganizing at object granularity, and is designed to improve these backends.

\myparagraph{Garbage Collection and Object Relocation}
GCs have long relocated objects to improve locality: generational collectors cluster recent allocations, profile-guided reordering~\cite{oor-oopsla-2004} groups hot objects during copying, and the Bookmarking Collector~\cite{bc-pldi05} avoids paging during collection. \SYSTEM{} differs in relocation safety: where ZGC~\cite{zgc} and Shenandoah~\cite{shenandoah} use load barriers to redirect accesses while threads hold stale pointers, \SYSTEM{} adopts a \emph{quiescence-based} approach inspired by epoch-based reclamation~\cite{fraser, faster} and RCU~\cite{rcu}, relocating only when an object's active-thread count reaches zero and aborting on any concurrent access via CAS failure. It also differs in purpose: GCs relocate to reclaim memory or improve cache locality, whereas \SYSTEM{} relocates to improve \emph{page utilization} so that tiering mechanisms see uniform pages.

\vspace{-1mm}
\section{Limitations and Applicability}
\label{sec:limitations}

\SYSTEM{} targets a specific class of code, pointer-based concurrent data structures in unmanaged languages, rather than aiming for universal applicability. Its requirements bound where it applies, which we make explicit so users can judge fit.

\begin{itemize}[leftmargin=*, topsep=2pt, itemsep=1pt]
\item \textbf{No pointer stability.} Relocation invalidates raw pointers cached across operations (e.g., addresses retained by Abseil maps or STL iterators). Callers must re-resolve through the guide, as pointer-unstable containers already require~\cite{abseil}.
\item \textbf{Single ownership.} A guide asserts exclusive ownership like \texttt{std::unique\_ptr}, and migration updates only that guide. Multiple guides aliasing one object (shared graph or doubly-linked nodes) are unsupported. CrestDB enforces this by deep-copying on insert.
\item \textbf{No pointer arithmetic.} Guides do not support \texttt{+}/\texttt{[]} across object boundaries, so contiguous arrays, matrices, and packed columnar data are not supported in \SYSTEM{}; a validation pass rejects such uses.
\item \textbf{Language support.} The design relies on dereference-operator overloading, which is not supported by Go or Java.
\item \textbf{Annotation.} Developers mark a small set of pointer fields, so adoption is not fully transparent; the compiler handles the rest (\S\ref{sec:llvm-pass}).
\item \textbf{Hardware composability.} The guide encoding reuses high-order address bits, which can contend with Intel LAM~\cite{lam}, ARM TBI~\cite{tbi}, or HWASAN~\cite{hwasan} if those features claim the same bits; wider addressing~\cite{128b-linux} relaxes this.\\
\end{itemize}
We target unmanaged languages, as most latency-sensitive datacenter code is C++ or Rust; the mechanisms generalize to managed runtimes.
\vspace{-1mm}
\section{Conclusion}

\SYSTEM{} shows that memory tiering improves when applications cooperate with the OS rather than bypass it.
By reorganizing virtual address space so that page boundaries align with object temperature, \SYSTEM{} makes existing backends more effective without modifying them.
The approach has limitations: it requires developer annotation of relocatable pointers and applies only to pointer-based structures.
However, the principle of address-space engineering extends beyond tiering.
The same techniques could improve generational garbage collection (grouping by access intensity rather than age), reduce false sharing of pages across NUMA nodes (separating objects by access pattern), and strengthen isolation (grouping by trust level).
An application's address-space layout is itself a channel to the OS~\cite{ArpaciDusseau23-Book}: by grouping objects of similar temperature, \SYSTEM{} turns layout into an object-level signal that unmodified page-based tiering can act on, with no new interface between the two.

\section*{Acknowledgments}
We thank David Culler, Abhinav Sharma, Lilian Tsai, Qian Ge, Teresa Johnson, Sujay Yadalam, colleagues in SystemsResearch@Google and the ADvanced Systems Laboratory (ADSL), the anonymous reviewers, and our shepherd for valuable feedback. This work was supported by gifts from Google.


\bibliographystyle{ACM-Reference-Format}
\bibliography{ref}
\end{document}